\documentstyle[12pt]{article}

\newcommand{\al}{\alpha}
\newcommand{\ep}{\epsilon}

\newcommand{\df}{\stackrel{\rm def}{=}}

\newcommand{\lb}{\lbrack}
\newcommand{\rb}{\rbrack}

\newcommand{\msc}[1]{\mbox{\scriptsize #1}}
\newcommand{\dsp}{\displaystyle}

\newcommand{\bc}{\mbox{{\bf C}}}
\newcommand{\br}{\mbox{{\bf R}}}
\newcommand{\bz}{\mbox{{\bf Z}}}

\newcommand{\bsz}{\msc{{\bf Z}}}

\newcommand{\cB}{{\cal B}}

\newcommand{\cG}{{\cal G}}
\newcommand{\cJ}{{\cal J}}

\newcommand{\cO}{{\cal O}}
\newcommand{\cM}{{\cal M}}
\newcommand{\cN}{{\cal N}}
\newcommand{\cD}{{\cal D}}
\newcommand{\cS}{{\cal S}}
\newcommand{\cP}{{\cal P}}
\newcommand{\cF}{{\cal F}}
\newcommand{\cQ}{{\cal Q}}
\newcommand{\cH}{{\cal H}}

\newcommand{\ket}[1]{{|#1\rangle}}

\newcommand{\Th}[2]{\Theta_{#1,#2}}
\newcommand{\th}{{\theta}}
\newcommand{\tTh}[2]{\tilde{\Theta}_{#1,#2}}
\newcommand{\ch}[2]{\mbox{ch}^{#1}_{#2}}

\newcommand{\Ch}[2]{\mbox{Ch}^{#1}_{#2}}

\newcommand{\tr}{\mbox{Tr}}
\newcommand{\mod}{\mbox{mod}}

\newcommand{\NS}{\mbox{NS}}
\newcommand{\tNS}{\widetilde{\mbox{NS}}}
\newcommand{\R}{\mbox{R}}
\newcommand{\tR}{\widetilde{\mbox{R}}}
\newcommand{\sNS}{\msc{NS}}
\newcommand{\stNS}{\widetilde{\msc{NS}}}
\newcommand{\sR}{\msc{R}}
\newcommand{\stR}{\widetilde{\msc{R}}}

\newcommand{\nn}{\nonumber\\}

\newcommand {\eqn}[1]{(\ref{#1})}

\makeatletter
\@addtoreset{equation}{section}
\def\theequation{\thesection.\arabic{equation}}
\makeatother

\begin{document}
\vskip 7mm

\begin{titlepage}
 
 \renewcommand{\thefootnote}{\fnsymbol{footnote}}
 \font\csc=cmcsc10 scaled\magstep1
 {\baselineskip=14pt
 \rightline{
 \vbox{\hbox{hep-th/0207124}
       \hbox{UT-02-38}
       }}}

 \vfill
 \baselineskip=20pt
 \begin{center}
 \centerline{\Huge  Superstring Vacua of 4-dimensional PP-Waves} 
 \vskip 5mm 
 \centerline{\Huge   with Enhanced Supersymmetry}

 \vskip 2.0 truecm
\noindent{\it \large Yasuaki Hikida and Yuji Sugawara} \\
{\sf hikida@hep-th.phys.s.u-tokyo.ac.jp~,~
sugawara@hep-th.phys.s.u-tokyo.ac.jp}
\bigskip

 \vskip .6 truecm
 {\baselineskip=15pt
 {\it Department of Physics,  Faculty of Science, \\
  University of Tokyo \\
  Hongo 7-3-1, Bunkyo-ku, Tokyo 113-0033, Japan}
 }
 \vskip .4 truecm

 \end{center}

 \vfill
 \vskip 0.5 truecm

\begin{abstract}
\baselineskip 6.7mm

We study the superstring vacua constructed from the conformal
field theories of the type $H_4 \times \cM$,
where $H_4$ denotes the super Nappi-Witten model 
(super WZW model on the 4-dimensional Heisenberg group $H_4$) 
and $\cM$ denotes an arbitrary $\cN=2$ rational superconformal 
field theory with $c=9$. We define (type II) superstring vacua with 8  
supercharges, which are twice as many as those on the backgrounds
of $H_4 \times CY_3$.
We explicitly construct as physical vertices the space-time SUSY algebra 
that is a natural extension of $H_4$ Lie algebra. 
The spectrum of physical states is classified into two sectors: 
(1) strings freely propagating along the transverse plane of pp-wave 
geometry and possessing the integral $U(1)_R$-charges in $\cM$ sector, 
and (2) strings that do not freely propagate  along the transverse
plane and possess the fractional $U(1)_R$-charges in $\cM$. 
The former behaves like  the string excitations in the usual 
Calabi-Yau compactification, but the latter defines new sectors without
changing the physics in ``bulk'' space. 
We also analyze the thermal partition functions of these systems, 
emphasizing the similarity to the DLCQ string theory. As a byproduct
we prove the supersymmetric cancellation of conformal blocks in 
an arbitrary unitary $\cN=2$ SCFT of $c=12$ with 
the suitable GSO projection.

\end{abstract}

\vfill

\setcounter{footnote}{0}
\renewcommand{\thefootnote}{\arabic{footnote}}
\end{titlepage}
\baselineskip 18pt

\newpage
\section{Introduction}

Four dimensional string vacua with unbroken space-time supersymmetry (SUSY)   
have been a subject of great importance for a manifest physical reason.
The most familiar examples of them  are described 
by the Calabi-Yau compactifications; $\br^{3,1}\times CY_3$.
A non-trivial extension to curved four dimensional space-times possessing
space-time SUSY has been given by considering the pp-wave backgrounds,
which admit light-like Killing vectors \cite{old pp,NW,KK,Sfetsos,RT2,NW2}.    
Among other things, Nappi-Witten (NW) model \cite{NW}, 
which is the WZW model based on the four dimensional 
Heisenberg group $H_4$ (or equivalently the central extension of
two dimensional Poincare group $E_2^c$), 
has been receiving many attentions \cite{KK,Sfetsos,RT2,NW2}
and possesses good properties to handle:
(1) this model has an exact world-sheet 
conformal symmetry to all orders of $\al'$ with a constant dilaton
and the central charge is precisely equal to 4 (6 for the supersymmetric
model), and 
(2) this model can be solved exactly by current algebra techniques,
since it is a WZW model.

In this paper we study the superstring vacua constructed from the
superconformal field theories of the type $H_4 \times \cM$,
where $H_4$ denotes the super NW model 
(super WZW model on $H_4$) \cite{KK}
and $\cM$ denotes an arbitrary $\cN=2$ unitary
rational superconformal field theory with $c=9$.
In order to define superstring vacua 
based on the RNS formalism we need to introduce
a consistent ``GSO projection'' that ensures 
the mutual locality of space-time supercharges as in the Gepner models 
\cite{Gepner}.
The simplest choice of the GSO projection is to restrict to
the sectors with integral $U(1)_R$-charges in $\cM$.
Such string vacua correspond to nothing but 
the background $H_4 \times CY_3$,
and have generically  4 supercharges (half of maximal SUSY).

More interesting choice of GSO projection is to impose the integrality
of {\em total\/} $U(1)_R$-charge in $H_4\times \cM$. 
This model is the primary concern in this paper and leads 
to a theory with enhanced SUSY,
that is, (at least) 8 supercharges 
(maximal SUSY for $CY_3$ compactification).
As observed in many cases of pp-wave models with enhanced SUSY
\cite{BFHP}, a large class of these 
string vacua can be reinterpreted as the Penrose limits of 
$AdS$ backgrounds. To be more accurate 
they can be constructed from the  $AdS_3 \times S^1 \times \cM'(k)$ 
background discussed in \cite{GRBL}, 
where the level of (bosonic) $SL(2;\br)$ current algebra is $k+2$ and 
the $\cM'(k)$ denotes an arbitrary $\cN=2$ rational superconformal
field theory which possesses one parameter $k$ such that  
$\dsp c=9-\frac{6}{k}$. 
The space-time energy corresponds  to $-J^3_0$ 
($J^A$ are the total currents of $SL(2;\br)$ describing
$AdS_3$ sector) and the space-time R-charge is measured by $K_0$
($K$ is the $U(1)$ current associated with the $S^1$ sector.)
The Penrose limit is expressed as $k \to \infty$ with keeping 
the value $\dsp \frac{1}{k}(K_0-J^3_0)$ finite and 
$|K_0+J^3_0|\ll k$ as discussed in \cite{Sfetsos}. 
This contraction of current algebra amounts to focusing on 
the almost BPS states with large $R$-charges as in \cite{BMN}. 
This point is another motivation of this work, and
it is  a natural extension of our previous
study of the Penrose limit of $AdS_3\times S^3 \times M^4$ \cite{HS}.

A limited list of recent works related to this paper is given in
\cite{BMN,GO,RT,KP,TT,PS,Das,Lunin,Michishita,Gomis}.

~

This paper is organized as follows: 
In section 2 we make a brief review on super Nappi-Witten model.
In section 3 we present two types of superstring vacua based on 
the conformal field theories $H_4 \times \cM$. One of them corresponds to the
background $H_4 \times CY_3$ and has 4 supercharges. The other type 
has 8 supercharges and can be regarded as the Penrose limit of 
$AdS_3 \times S^1 \times \cM'$ backgrounds. We also study in detail 
the spectrum of physical states in this model.  
In section 4 we compute the one-loop partition functions 
in order to check the consistency of the proposed string vacua.
Although the transverse sector contains conformal blocks possessing
good modular properties, the calculation in the longitudinal sector
leads to a difficulty of divergence.
We thus evaluate the partition functions of the thermal models 
to avoid this problem.
Section 5 is devoted to present a summary and some discussions.

~


\section{Super Nappi-Witten Model} 

We start with giving a short review on super Nappi-Witten (NW) model.
This model is defined as the super WZW model on
the four dimensional Heisenberg group $H_4$. 
It is described by the following supercurrents; 
\begin{eqnarray}
&& {\cal J} (\theta , z) = \psi_J (z) + \theta J (z) ~,~~
 {\cal F} (\theta , z) = \psi_F (z) + \theta F (z) ~, \nn
&& {\cal P} (\theta , z) = \psi_{P} (z) + \theta P(z) ~,~~
 {\cal P}^* (\theta , z) = \psi_{P^*} (z) + \theta P^* (z) ~.
\end{eqnarray}
The ``total currents''  $J(z)$, $F(z)$, $P(z)$ and $P^*(z)$ 
satisfy the OPEs
\begin{eqnarray}
 J(z) P(w) &\sim& \frac{P(w)}{z-w} ~,~~
 J(z) P^*(w) ~\sim~ - \frac{P^*(w)}{z-w} ~, \nn
 P(z) P^* (w) &\sim&  
 \frac{1}{(z-w)^2} + \frac{F(w)}{z-w}~,\nn
 J (z) F (w) &\sim& \frac{1}{(z-w)^2}~.
\label{H4OPE}
\end{eqnarray}
Other OPEs have no singular terms. 
Their superpartners are defined by the OPEs
\begin{eqnarray}
&&  \psi_{P} (z) \psi_{P^*} (w) \sim \frac{1}{z-w} ~,~~
 \psi_J (z) \psi_F (w) \sim \frac{1}{z-w} ~,\nn
&& J (z) \psi_{P} (w) \sim \psi_J(z)P(w) \sim
\frac{\psi_{P}(w)}{z-w} ~,\nn
&& J (z) \psi_{P^*} (w) \sim \psi_J(z)P^*(w) \sim
- \frac{\psi_{P^*}(w)}{z-w} ~,\nn
&& P (z) \psi_{P^*} (w) \sim \psi_{P}(z)P^*(w) \sim
\frac{\psi_{F}(w)}{z-w} ~.
\end{eqnarray}

We can employ the free field representations 
of the supercurrent algebra as given in \cite{KK};
\begin{eqnarray}
&& J=i\partial X^-~, ~~~ F=i\partial X^+ ~, \nn
&& P=(i\partial Z+ \psi^+\psi)e^{iX^+} ~, ~~~
P^*=(i\partial Z^*- \psi^+\psi^*)e^{-iX^+} ~,\nn
&& \psi_F = \psi^+~,~~~ \psi_J=\psi^-~, ~~~ 
\psi_P = \psi e^{iX^+}~,~~~ \psi_{P^*} = \psi^* e^{-iX^+} ~,
\end{eqnarray}
where the free fields $X^{\pm}$, $Z$, $Z^*$, $\psi^{\pm}$, $\psi$, 
and $\psi^*$ are defined by
\begin{eqnarray}
&& i\partial X^+(z) i\partial X^-(w) \sim \frac{1}{(z-w)^2} ~, ~~~
i\partial Z (z) i\partial Z^*(w) \sim \frac{1}{(z-w)^2} ~, \nn
&& \psi^+(z) \psi^-(w) \sim  \frac{1}{z-w} ~, ~~~
\psi(z)\psi^*(w) \sim \frac{1}{z-w}~. 
\end{eqnarray}
We actually have  the extended $\cN=2$ superconformal symmetry,
which is described most easily by these free fields as
\begin{eqnarray}
&& T_{H_4}(z) = -\partial X^+ \partial X^- - \partial Z \partial Z^* 
-\frac{1}{2} (\psi^+ \partial \psi^- -\partial \psi^+ \psi^-)
-\frac{1}{2} (\psi \partial \psi^* -\partial \psi  \psi^*) ~,\nn
&&G^+_{H_4}(z)=\psi^+i\partial X^- + \psi i\partial Z^* ~, ~~~
G^-_{H_4}(z)=\psi^-i\partial X^+ + \psi^* i\partial Z ~,\nn
&& I_{H_4}(z) = \psi^+\psi^- + \psi \psi^* ~.
\label{N=2 H4}
\end{eqnarray}
They actually generate an $\cN=2$ superconformal algebra (SCA)
with $c=6$\footnote
    {To be more precise the world-sheet superconformal symmetry 
     can be extended
      to $\cN=4$ according to the well-known properties of $\cN=2$
     SCFT, since the central charge is now equal to 6. The explicit 
     realization of $\cN=4$ SCA in super NW model is given in
     \cite{KK}. However, the $\cN=4$ structure is not necessary 
      in the analysis of this paper.}.
We also note that the ``transverse coordinates'' 
$\{Z,\,Z^*,\,\psi,\,\psi^*\}$ generate an $\cN=2$ SCA 
with $c=3$ in a manifest way.


The irreducible representations of NW current algebra are classified 
in \cite{KK}. (See also \cite{KP,HS}.) Only the non-trivial point
is the existence of spectral flow symmetry;
\begin{eqnarray}
 J_n \to J_n ~,~~ F_n \to F_n + p \delta_{n,0} ~,~~ 
 P_{n} \to P_{n + p} ~,~~P_{n}^* \to P_{n-p}^* ~, \nn
 \psi_{J,\,r} \to \psi_{J, \, r} ~,~~ \psi_{F,\,r} \to \psi_{F,\,r} ~,~~ 
 \psi_{P,\,r} \to \psi_{P,\, r + p} ~,~~
 \psi_{P^*,\,r} \to \psi_{P^*,\,r-p}~.
\label{spectral flow}
\end{eqnarray}
The (spectrally flowed) type II representations are defined as in \cite{HS};
\begin{eqnarray}
&& J_0 \ket{j,\eta, p} = j \ket{j,\eta, p} ~,~~
 F_0 \ket{j,\eta, p} =  (\eta+p)  \ket{j,\eta, p} ~,\nn
&& P_{n} \ket{j,\eta,p} = 0 ~,~({}^\forall n \geq -p) ~,~~
 P_{n}^* \ket{j,\eta,p} = 0 ~,~({}^\forall n > p) ~, \nn
&& \psi_{J,\,r} \ket{j,\eta,p} =
  0 ~,~({}^\forall r  > 0)~ ,~~
 \psi_{F,\,r} \ket{j,\eta,p} = 
  0 ~,~({}^\forall r  > 0)~ , \nn
&& \psi_{P,\,r} \ket{j,\eta,p} = 
  0 ~,~({}^\forall r  > -p)~ ,~~
  \psi_{P^*,\,r} \ket{j,\eta,p} = 
  0 ~,~({}^\forall r  > p)~ ,
\label{flowed type II}
\end{eqnarray}
where $r \in 1/2 + \bz$. The type III representations are
similarly defined for $-1<\eta<0$. The type I representations
correspond to the case of $\eta=0$, in which we have no highest weight 
and lowest weight states (for $p=0$) and have extra zero-mode momenta
$P_0$ and  $P^*_0$.
In terms of free fields the vacuum vector of \eqn{flowed type II} 
corresponds to the next vertex operator
\begin{equation}
e^{ijX^+ + i(p+\eta)X^-}\sigma_{\eta}~,
\end{equation}
where $\sigma_{\eta}$ denotes the twist field defined by
\begin{eqnarray}
&& i \partial Z (z) \sigma_\eta (w) \sim 
 (z-w)^{- \eta} {\tau}_\eta (w) ~,~~
 i \partial Z^* (z) \sigma_\eta (w) \sim 
 (z-w)^{\eta - 1} {\tau '}_\eta (w) ~, \nn
&& \psi (z) \sigma_\eta (w) \sim 
 (z-w)^{- \eta} t_\eta (w) ~,~~
 \psi^* (z) \sigma_\eta (w) \sim 
 (z-w)^{\eta} t'_\eta (w) ~. 
\end{eqnarray}
This twist field 
$\sigma_{\eta}$ has the conformal weight
\begin{equation}
h(\sigma_{\eta}) = \frac{1}{2}\eta(1-\eta) + \frac{1}{2}\eta^2 
= \frac{1}{2} \eta~,
\end{equation}
and $U(1)_R$-charge 
\begin{equation}
Q(\sigma_{\eta}) = -\eta~.
\end{equation}

The type I representations reduce to the usual Fock representations
of free oscillators with no twists. The Fock vacua are expressed as 
\begin{equation}
e^{i(j+n)X^+ + ipX^-+iq^*Z+iqZ^*}~,~~~({}^{\forall} n\in \bz)~.
\end{equation}
This sector corresponds to the string modes freely propagating 
along the transverse plane $Z$ and $Z^*$.

~


\section{Superstring Vacua based on $H_4 \times \cM$}

Now we study the superstring vacua constructed from the conformal theory 
$H_4 \times \cM$,
where we assume $\cM$ is an $\cN=2$ unitary rational 
SCFT with $c=9$. The most non-trivial task is to introduce the GSO 
condition compatible with unbroken space-time SUSY. 

We begin with clarifying the bosonic symmetry algebra.
In the covariant gauge the physical vertices are characterized by 
the BRST charge that has the standard form 
\begin{equation}
Q_{\msc{BRST}} = \oint \left\lb c\left(T-\frac{1}{2}(\partial \phi)^2
  -\partial^2\phi -\eta\partial\xi + \partial c b\right)
+\eta e^{\phi}G-b\eta\partial\eta e^{2\phi}\right\rb  ,  \\
\end{equation}
where $T\equiv T_{H_4}+T_{\cM}$ and $G\equiv G^+_{H_4}+G^-_{H_4}
+G^+_{\cM}+G^-_{\cM}$ 
are the total stress tensor and the superconformal current, respectively.
We also introduced the standard ghost system $(b,c)$ and  bosonized 
superghost system $(\phi, \xi, \eta)$.
The bosonic symmetry algebra 
consists of the zero-modes of total $H_4$ currents,
which are manifestly BRST invariant;
\begin{eqnarray}
&& \cJ = \oint \psi^- e^{-\phi} 
    =  \oint i \partial X^- =J_0  ~, \nn
&& \cF = \oint \psi^+ e^{-\phi}   
     =   \oint i \partial X^+ =F_0 ~,\nn
&& \cP = \oint \psi e^{i  X^+} e^{-\phi}  
= \oint \left( i \partial Z + \psi^+\psi \right)e^{i X^+}= P_{0}~,\nn
&&  \cP^* ~=~ \oint \psi^* e^{- i  X^+} e^{-\phi} 
= \oint\left( i \partial Z^* 
- \psi^+\psi^* \right) e^{- i  X^+} = P_{0}^*  ~.
\label{global bosonic}
\end{eqnarray}
These operators generate the $H_4$ Lie algebra as expected;
\begin{eqnarray}
&& \lb \cJ, \cP \rb = \cP~, ~~~ \lb \cJ, \cP^* \rb=-\cP^* ~, ~~~
 \lb \cP, \cP^* \rb = \cF ~.
\label{bosonic symmetry}
\end{eqnarray}

The Fock vacua are  characterized by the eigen-values of $\cF$,
$\bar{\cF}$, $\cJ$ and $\bar{\cJ}$. It is here important to point 
out the fact that our free fields $X^{\pm}$, $Z$ and  $Z^*$
are not the sigma model coordinates as clarified in \cite{KP}.
In particular, the left and right movers of $X^+$ are
regarded as those defined with respect
to the same coordinate system, but those of $X^-$ are not.
We hence have to assume 
\begin{equation}
\cF=\bar{\cF}=p+\eta~,~~~(p\in \bz,~0\leq \eta<1) ~,
\label{F}
\end{equation} 
unless considering an additional compactification.
However, it is possible to have ``helicitiy in the transverse plane''
\cite{KP} along $X^-$ direction;
\begin{equation}
\cJ-\bar{\cJ}=h\in \bz~,
\label{J}
\end{equation}
which will play an important role in our later discussion.


In order to describe the space-time SUSY  we must introduce
the spin fields (up to cocycle factors) 
\begin{equation}
S^{\ep_0\ep_1\ep_2} = 
e^{i\left(\frac{\ep_0}{2}H_0+\frac{\ep_1}{2}H_1
+\frac{\ep_2}{2}\sqrt{3}H_2\right)}~,
\label{spin fields}
\end{equation}
where $\ep_i = \pm $.  
The free scalars $H_i$ are defined by
\begin{equation}
i\partial H_0=\psi^+\psi^-~,~~~ i\partial H_1 = \psi\psi^*~,~~~
\sqrt{3}i \partial H_2 = -I_{\cM}~,
\end{equation}
where $I_{\cM}$ denotes the $U(1)_R$-current in the $\cM$ sector and 
satisfies the OPE
\begin{equation}
I_{\cM}(z)I_{\cM}(w) \sim \frac{3}{(z-w)^2}~.
\end{equation}
In order to define superstring vacua
we must enforce the GSO condition which assures
the locality of the supercharges. 
We shall consider the following two cases, which will be analysed
separately.

~

\subsection{Superstring Vacua of $H_4\times CY_3$}

We first consider the simpler case. We impose as the GSO
condition
\begin{equation}
  I_{\cM,\,0} \in \bz ~.
\label{GSO 0}
\end{equation}
This condition converts the SCFT $\cM$ into a $\sigma$-model on $CY_3$
(in a broad sense) as in the Gepner models.  Therefore, 
we obtain the background $H_4\times CY_3$,
which is a naive extension of familiar string vacua $\br^{3,1}\times CY_3$.

Precisely speaking, one must of course further enforce the standard
GSO projection with respect to the spin structures. 
For the spin fields we obtain
\begin{equation}
\prod_{i=0}^2 \ep_i = +1~,
\end{equation} 
in our convention.

The space-time supercharges are explicitly constructed as follows
\begin{equation}
\cQ^{\pm} =\oint S^{+\pm\pm}e^{\pm  \frac{i}{2} X^+}e^{-\frac{\phi}{2}}~,
\label{SUSY charge 0}
\end{equation}
(and their counter parts in the right mover).
They are obviously BRST invariant 
and act locally on all the states constrained by the above 
GSO condition\footnote
   {To be more precise, we must restrict  $p\in 2\bz$ for the
locality of \eqn{SUSY charge 0}. Nevertheless, we can incorporate 
all the representations of $H_4$ current algebra into the physical 
Hilbert space owing to the equivalence between 
the type II and type III representations by spectral flow;
$$
 \cH^{(II)}_{p, \eta} \cong \cH^{(III)}_{p+1,\eta-1}~,
$$
as we will again note in the later discussions. 
}.

It is easy to show that \eqn{SUSY charge 0} are in fact the totality
of possible supercharges, namely, the spin fields of the type $S^{-**}$
are not allowed. In fact,
the BRST invariance and the mutual locality are not
compatible for these operators. 
We hence (generically)  obtain 4 supercharges, 
and this fact is consistent with the analysis 
of Killing spinors in type II supergravity on generic pp-wave
backgrounds (see, for example, \cite{BKO}),
in which the half chirality should be projected out.

The SUSY algebra is quite simple;
\begin{eqnarray}
&& \{\cQ^+,\cQ^-\} = \cF~, \nn
&& \{\cQ^{\pm}, \cQ^{\pm}\}=0~,\nn
&& \lb \cJ, \cQ^{\pm}\rb = \pm \frac{1}{2}\cQ^{\pm}~,
\label{SUSY alg 0}
\end{eqnarray}
and other combinations of commutation relations with
$\cF, \cJ, \cP$ and $\cP^*$ vanish. 

~

A few comments are in order:

~

\noindent
{\bf 1.} Since the GSO projection \eqn{GSO 0} acts solely on 
$\cM$, we can freely choose the twist parameter $\eta$. 
This aspect is in a sharp contrast with the string vacua with
enhanced SUSY we will discuss below.

~

\noindent
{\bf 2.} Both of supercharges $\cQ^+$, $\cQ^-$ \eqn{SUSY charge 0}
do not commute with the light-cone Hamiltonian 
$H_{\msc{l.c.}} \df -(\cJ+\bar{\cJ}) $, which is essentially
the transverse Virasoro operator because of the on-shell condition.
This fact implies that the number of physical states in the NS and R sectors
is {\em not\/} balanced at each energy level. 
It sounds peculiar since we are now considering  supersymmetric vacua in which 
Killing spinors exist.
However, we can also employ the different light-cone Hamiltonian 
based on the quantization in the different coordinate system,
which is related to the above $H_{\msc{l.c.}}$ by a shift of 
``angular momentum operator'' \cite{RT2,RT}. 
The right-moving supercharges $\bar{\cQ}^{\pm}$ do not commute with 
that Hamiltonian, while the left-movers $\cQ^{\pm}$ commute.
The supersymmetric cancellation for these vacua is realized in this
sense.

~

\subsection{Superstring Vacua with Enhanced Supersymmetry}

Let us next consider a different choice of GSO condition.
This is the main subject in this paper. 
We shall take
\begin{equation}
  I_{\msc{tot},0}\equiv I_{H_4,\,0} + I_{\cM,\,0} \in \bz~,
\label{GSO}
\end{equation}
where $I_{H_4}$ denotes the $U(1)_R$-current in the $H_4$ sector 
introduced in \eqn{N=2 H4}. 
The GSO condition incorporating spin structures further projects out
the half degrees of freedom as usual.  Especially, we obtain
\begin{equation}
  I_{\msc{tot},0} \in 2\bz+1~,
\label{GSO NS}
\end{equation}
for the NS sector.

The non-trivial difference from \eqn{GSO 0} 
is the existence of extra contribution to the $U(1)_R$-charges from
the twist field $\sigma_{\eta}$. It leads to a different locality
condition of supercharges, and we find that the proper
supercharges are 
\begin{eqnarray}
&& \cQ^{\pm\pm} = \oint S^{+\pm\pm}e^{\pm i X^+}e^{-\frac{\phi}{2}}~,
\label{SUSY charge 1}\\
&& \cQ^{\pm\mp} = \oint S^{-\pm\mp}e^{-\frac{\phi}{2}}~,
\label{SUSY charge 2}
\end{eqnarray}
and their counterparts in the right mover.
We thus obtain 8 supercharges enhanced  
twice  compared with the previous case $H_4\times CY_3$.
These operators generate the following SUSY algebra together with
\eqn{bosonic symmetry};
\begin{eqnarray}
&& \lb \cJ, \cQ^{\pm\pm}\rb = \pm \cQ^{\pm\pm} ~, ~~~
\lb \cJ, \cQ^{\pm\mp} \rb = 0~,\nn
&& \lb \cQ^{-+}, P \rb = \cQ^{++}~,~~~ \lb \cQ^{+-}, P^*\rb =-\cQ^{--}~, \nn
&& \{\cQ^{++}, \cQ^{--}\} = - \cF~,~~~ \{\cQ^{+-}, \cQ^{-+}\}=\cJ~, \nn
&& \{\cQ^{+-}, \cQ^{++}\} = P ~, ~~~ \{\cQ^{-+}, \cQ^{--}\}= P^*~.
\label{SUSY pp-wave}
\end{eqnarray}
This is a natural supersymmetric extension of 
$H_4$ Lie algebra and can be derived by 
contracting the ``zero-mode subalgebra''
$\{L_0, L_{\pm 1}, I_0, G^{\pm}_{\pm 1/2}, G^{\pm}_{\mp 1/2}\}$
of $\cN=2$ superconformal algebra. This aspect of course reflects 
the fact that the string vacua  of this type can be obtained 
as the Penrose limit of the  $AdS_3\times S^1 \times \cM'(k)$
superstring \cite{GRBL}, 
as we already mentioned.

~

We here make a few comments:

~

\noindent
{\bf 1.} Because we are now choosing the GSO condition \eqn{GSO} 
rather than \eqn{GSO 0}, the possible value of twist parameter 
$\eta$ should depend on the spectrum of $\cM$ sector. 
In particular, since $\cM$ is assumed to be a rational SCFT,
only the rational values of $\eta$ are allowed. This fact 
makes the total conformal system easier to deal with 
from the view points of modular invariance.

~

\noindent
{\bf 2.} We now have the supercharges including the spin fields of 
the type $S^{-**}$ in contrast to the previous case. Such supercharges,
that is, $\cQ^{\pm \mp}$, commute with $\cJ$, 
implying  that the physical states in the NS and R
sectors are manifestly balanced at each energy level.

~

Now, let us analyse the physical Hilbert space for each representations.

~

\noindent{\bf 1. Type I representations}

For the (flowed) type I representations things become very easy, since
we have no twisted coordinates. 
In this case the GSO condition \eqn{GSO} allows 
the physical states only in the integral $U(1)_R$-charge sectors of $\cM$.
Hence it seems that the physical spectrum simply reduces to 
that of $\br^{3,1}\times \cM|_{U(1)\msc{-projected}} 
\cong \br^{3,1} \times CY_3$.

However, there is  a slight difference. 
Since the spectral flow parameter $p$ is discrete, we should have
the discretized light-cone momentum
\begin{equation}
\cF=\bar{\cF}=p\in \bz~.
\end{equation}
Moreover, the on-shell condition and \eqn{J} yield the following
level matching condition
\begin{eqnarray}
L_0^{\msc{tr}}-\bar{L}_0^{\msc{tr}} \in p\bz~.
\end{eqnarray} 
Therefore  we have obtained the physical spectrum which is the same as
that of the DLCQ (discrete light-cone quantization) superstring theory
\cite{DLCQ}
on $\br^{3,1} \times CY_3$ with the null compactification
$X^- \sim X^- + 2\pi$.

~

\noindent{\bf 2. Type II (and type III) representations}

The sectors including type II and type III representations 
are more interesting and contain new physical states that are absent 
in the usual Calabi-Yau compactifications. Strings in these sectors
cannot freely propagate along the transverse directions 
(in the four dimensional space-time), because these sectors are 
described by the twisted string coordinates that have no zero-modes.

It is especially important to study the BPS states. 
We only focus on the type II representations $(0<\eta<1)$ and 
the analysis is parallel for the type III representations $(-1<\eta<0)$.
The BPS states are characterized by the condition $\cJ=0$. We thus
start with the candidates of the forms (in the NS sector)
\begin{eqnarray}
&& e^{i(p+\eta)X^-}\sigma_{\eta}\ket{0}_{H_4}\otimes\cO\ket{0}_{\cM}\otimes 
ce^{-\phi}\ket{0}_{\msc{gh}} ~, \label{BPS 1}\\
&& \psi_{-1/2+\eta}e^{i(p+\eta)X^-}\sigma_{\eta}\ket{0}_{H_4}
\otimes \cO\ket{0}_{\cM}\otimes 
ce^{-\phi}\ket{0}_{\msc{gh}} ~, \label{BPS 2}
\end{eqnarray}
where $\cO$ denotes an arbitrary (anti) chiral primary field in the $\cM$
sector which has $U(1)_R$-charge $Q(\cO)$ (and hence the conformal weight
$\dsp h(\cO)=\frac{1}{2}Q(\cO)$ for chiral primary and 
$\dsp h(\cO)=-\frac{1}{2}Q(\cO)$ for anti-chiral primary).
For the ``tachyon like'' state \eqn{BPS 1} 
the on-shell condition gives us 
\begin{equation}
     \eta + |Q(\cO)| =1~.
\label{on-shell BPS 1}
\end{equation}
The GSO condition \eqn{GSO NS} can be written as  
\begin{equation}
     -\eta + Q(\cO) \in 1+ 2\bz~,
\end{equation} 
which is compatible with \eqn{on-shell BPS 1} in the case of anti-chiral
primary states $Q(\cO)<0$.

On the other hand, for the ``graviton like'' state \eqn{BPS 2} 
the on-shell condition leads to 
\begin{equation}
     \eta = |Q(\cO)| ~,
\label{on-shell BPS 2}
\end{equation}
which is compatible with the GSO condition
\begin{equation}
     -\eta + Q(\cO) \in 2\bz~,
\end{equation} 
in the case of chiral primary states $Q(\cO)>0$.

We remark that the states with fractional $Q(\cO)$ contribute 
to these sectors. Such physical states are absent
in the usual Calabi-Yau compactifications. Moreover, we should note the 
fact that all the world-sheet (anti) chiral primaries $\cO$ do not necessarily
correspond to the space-time BPS states. In fact, we could use  
the (anti) chiral primaries 
with $|Q(\cO)| \leq 3 $ in principle.
However, the on-shell conditions for the BPS states 
\eqn{on-shell BPS 2} cannot be always satisfied, since 
we have the constraint $0<\eta<1$.

More general physical states are created by the DDF operators, 
which are BRST invariant and act locally on the Fock vacua 
constrained by the GSO condition \eqn{GSO};
\begin{eqnarray}
&& \cP_n = \frac{1}{\sqrt{p+\eta}}
\oint \psi e^{i\frac{n+\eta}{p+\eta}X^+}e^{-\phi}~,\nn
&&\cP^*_n = \frac{1}{\sqrt{p+\eta}}
\oint \psi^* e^{i\frac{n-\eta}{p+\eta}X^+}e^{-\phi}~,\nn
&&\cQ^{\pm\pm}_n = \frac{1}{\sqrt{p+\eta}} \oint S^{+\pm\pm}
           e^{\frac{n\pm\eta}{p+\eta}X^+} e^{-\frac{\phi}{2}}~.
\label{DDF}
\end{eqnarray}
Obviously,  $\sqrt{p+\eta}\cP_p \equiv \cP$, 
$\sqrt{p+\eta}\cP^*_{-p} \equiv \cP^*$, 
$\sqrt{p+\eta}\cQ^{++}_p \equiv \cQ^{++}$,
$\sqrt{p+\eta}\cQ^{--}_{-p} \equiv \cQ^{--}$ and  
they satisfy the following (anti-)commutation relations
\begin{eqnarray}
&& \lb \cP_m, \cP^*_n \rb = \frac{m+\eta}{p+\eta}\delta_{m+n,0}~,~~~
 \{ \cQ^{++}_m, \cQ^{--}_n \} = - \delta_{m+n,0}~, \nn
&& \lb \cJ, \cP_n \rb = \frac{n+\eta}{p+\eta} \cP_n ~,~~~
 \lb \cJ, \cP^*_n \rb = \frac{n-\eta}{p+\eta} \cP^*_n ~, \nn
&& \lb \cJ, \cQ^{++}_n \rb = \frac{n+\eta}{p+\eta} \cQ^{++}_n ~,~~~
 \lb \cJ, \cQ^{--}_n \rb = \frac{n-\eta}{p+\eta} \cQ^{--}_n ~.
\label{DDF algebra 1}
\end{eqnarray}
Furthermore, $(\cP_n, \cQ^{++}_n)$ and  $(\cP_n^*, \cQ^{--}_n)$ 
compose supermultiplets with respect to $\cQ^{+-}$, $\cQ^{-+}$, namely,
\begin{eqnarray}
&& \lb \cQ^{-+}, \cP_n \rb = \frac{n+\eta}{p+\eta}\cQ^{++}_n~, ~~~
\{\cQ^{+-}, \cQ^{++}_n\} = \cP_n~, \nn
&& \lb \cQ^{+-}, \cP^*_n \rb = \frac{n-\eta}{p+\eta} \cQ^{--}_n~, ~~~
\{\cQ^{-+}, \cQ^{--}_n\} = \cP^*_n~.
\label{DDF algebra 2}
\end{eqnarray}

Unfortunately, these DDF operators alone cannot generate the full BRST 
cohomology in contrast to the case of Penrose limit of 
$AdS_3\times S^3\times M^4$ discussed in \cite{HS}.
This fact is due to the lack of detailed information of $\cM$ sector,
and it seems difficult to construct the concrete DDF operators describing 
the excitations in $\cM$ sector. However, we can nevertheless
analyze these excitations at least in the light-cone gauge quantization. 
We later compute  the one-loop partition functions,  which should 
contain the full information of physical states.

~

To close this section we present a brief discussion about the
$AdS_3/CFT_2$ correspondence. 
As we already mentioned, in some cases 
$H_4\times \cM$ models can be regarded as the Penrose limits of 
$AdS_3\times S^1 \times \cM'$, 
which amount to focusing on the almost BPS states with 
large $R$-charges as in \cite{BMN}. 
Until now, the satisfactory holographic dual theories are not known 
for the general string vacua of this type. 
However, some symmetric orbifold theories 
are proposed in \cite{AGS} as in $AdS_3 \times S^3 \times
M^4$ ($M^4= T^4$ or $K3$). 
In this sense it may be interesting to investigate to what extent 
we can relate the string spectrum in the $H_4 \times \cM$ models with
the almost BPS spectrum of some symmetric orbifold theories 
as a natural extension of our previous work \cite{HS}.

For arbitrary $\cM$ the story seems to be difficult. 
In fact, all the string vacua are not necessarily obtained 
from the vacua of the type $AdS_3 \times S^1 \times \cM'$. 
We thus specialize to the cases of $\cM' = M_k\otimes \cM_0$, 
where $M_k$ denotes the $(k+2)$-th $\cN=2$
minimal model ($c=3-6/k$) and $\cM_0$ is an arbitrary $\cN=2$ unitary 
rational CFT with $c=6$. We also assume that $k$ is equal to 
the level of $SL(2;\br)$ WZW model describing 
the $AdS_3$ string. In those cases the Penrose limits
$k\,\rightarrow\, \infty$ of $AdS_3 \times S^1 \times \cM'$
are described by the string vacua of the type we are now discussing.
In the cases $\cM_0=T^4$ or $T^4/\bz_2$ the model reduces 
to the simpler one $H_6 \times \cM_0$ which we studied in the previous
paper \cite{HS}. Therefore, it seems natural to expect 
that the dual theory of the present model is 
the superconformal theory of the type 
$Sym^M(\cM_0)\equiv \cM_0^M/S_M$, where $S_M$ means the $M$-th
symmetric group.

Let us now observe whether this proposal is correct.
We concentrate on the $\bz_N$-twisted sector, 
which corresponds to the single particle Hilbert space of the 
``long string of length $N$'' and is described by an $\cN=2$
SCA $\{\hat{L}_n,  \hat{I}_n,  \hat{G}^{\pm}_{r}\}$ with $c=6N$. 
The analysis of BPS states similar to \cite{HS} leads to the spectrum;
\begin{equation} 
h \equiv  \frac{Q}{2}= \frac{Q_{\cM_0}}{2} + \frac{1}{2}(N-1)~,
\label{BPS symmetric}
\end{equation}
where $Q_{\cM_0}$ denotes the possible $R$-charge of chiral primary fields
of $\cM_0$ ($0\leq Q_{\cM_0} \leq 2$). 
The Penrose limit corresponds to the large $N$, and in that case
the BPS states have approximately degenerate $R$-charge $Q\approx N$.
We should employ the identifications as in \cite{HS};
\begin{eqnarray}
&&
\cF ~\longleftrightarrow~ 
\frac{1}{k}\left(\frac{1}{2}\hat{I}_0+\hat{L}_0\right) ~,~~~
\cJ ~\longleftrightarrow~ 
\frac{1}{2}\hat{I}_0-\hat{L}_0 ~.
\label{FJ}
\end{eqnarray}
We thus  assign
\begin{equation}
N=k(p+\eta)~ .
\label{Nk}
\end{equation}
We can suppose that the $U(1)_R$-charges in  $\cM'\equiv M_k \otimes \cM_0$
are  quantized by the unit $1/k$ for sufficiently large $k$, 
and hence $\eta$ is also quantized as $\eta = l/k$ by the GSO condition
\eqn{GSO}. Therefore, \eqn{Nk} is a consistent relation and we can 
uniquely define $p$, $\eta$ from $N$, $k$.
In other words, we should define the ``Penrose limit'' of the present
symmetric orbifold theory as $N\rightarrow \infty$, $k\rightarrow
\infty$ with keeping $\cF=N/k$, $\cJ$ (under the identification
\eqn{FJ}) fixed to finite values.
Under this limit the $\cN=2$ SCA reduces to a super pp-wave algebra
\eqn{DDF algebra 1} and \eqn{DDF algebra 2}.
We can explicitly define the operators generating it as
\begin{eqnarray}
&& \cJ=\frac{1}{2}\hat{I}_0-\hat{L}_0~,~~~ 
\cF=\frac{1}{k}\left(\frac{1}{2}\hat{I}_0+\hat{L}_0\right)~,\nn
&& \cP_n=-\frac{1}{\sqrt{N}}
\left\{\frac{p+\eta}{n+\eta}\hat{L}_{\frac{n+\eta}{p+\eta}}
-\frac{1}{2}\left(\frac{p+\eta}{n+\eta}-1\right)
\hat{I}_{\frac{n+\eta}{p+\eta}}\right\}~,\nn
&& \cP^{*}_n=\frac{1}{\sqrt{N}}
\left\{\frac{p+\eta}{n-\eta}\hat{L}_{\frac{n-\eta}{p+\eta}}
-\frac{1}{2}\left(\frac{p+\eta}{n-\eta}+1\right)
\hat{I}_{\frac{n-\eta}{p+\eta}}\right\}~,\nn
&& \cQ^{\mp \pm}=\pm \hat{G}^{\pm}_{\mp \frac{1}{2}}~, ~~~
\cQ^{\pm\pm}_n = \frac{1}{\sqrt{N}} \frac{p+\eta}{n\pm \eta}
\hat{G}^{\pm}_{\frac{n\pm \eta}{p+\eta}\mp \frac{1}{2}}~.
\end{eqnarray}

At first glance this result seems to be satisfactory. 
However, we have a serious puzzle. 
In contrast to the $AdS_3\times S^3 \times M^4$ case,
all the chiral primaries in the internal CFT $\cM$ do {\em not\/}
necessarily appear in the spectrum of space-time BPS states. 
In fact, the chiral primaries with $1< Q_{\cM_0}\leq 2$ cannot define 
the BPS states in the pp-wave string spectrum, as we observed
before\footnote{One can find that 
    such missing BPS states correspond to non-normalizable
    states in the original $AdS_3 \times S^1 \times \cM'$ string theory
    before taking the Penrose limit,
    and hence do not appear in the physical Hilbert space. 
    This problem is supposed to originate from this fact.}.
On the other hand, we are not likely to have any reason to restrict 
the chiral primaries to the ones with $Q_{\cM_0}\leq 1$ 
in the symmetric orbifold theory.
We need make further investigation in order to understand completely
the aspects of holographic duality in these background\footnote
    {The holographic duality in the string vacua including the $H_4$ WZW 
     model has been also discussed in \cite{KP} in a different context. 
    It seems interesting to investigate the relation 
    between the duality proposed 
     in \cite{KP} and that originating from the 
     $AdS_3 \times S^1 \times \cM'$ background.}.

~

\section{One-Loop Partition Functions}

In this section we compute the one-loop  partition functions
in the string vacua with enhanced SUSY discussed in the previous
section. 
The partition function generally  has the following form
\begin{eqnarray}
&& Z_{\msc{1-loop}}
=\int_{\cF}\frac{d^2\tau}{\tau_2}\,\int\cD\lb X^+,X^-,\psi^+,\psi^-\rb
\cD\lb \mbox{gh}\rb\, e^{-S_{\msc{L}}-S_{\msc{gh}}}\,  
\tr\left((-1)^{\msc{\bf F}}q^{L_0^{\msc{tr}}-\frac{1}{2}}
\bar{q}^{\bar{L}_0^{\msc{tr}}-\frac{1}{2}}\right)~,
\label{partition 0}
\end{eqnarray}
where $S_{\msc{L}}$ and $S_{\msc{gh}}$ denote the actions of longitudinal
sector $\{X^+,X^-,\psi^+,\psi^-\}$ and the ghost sector, respectively.
Moreover,
$L_0^{\msc{tr}}$ and  $\bar{L}_0^{\msc{tr}}$ are the total Virasoro
operators of transverse sector $\{Z,Z^*,\psi,\psi^*\} \times \cM$
($c=12$), and $\mbox{\bf F}$ denotes the space-time fermion number (mod 2).
$\cF$ denotes the conventional fundamental domain of the moduli space
of torus;
\begin{equation}
\cF = \left\{\tau=\tau_1+i\tau_2\in \bc~;~
|\tau_1|\leq \frac{1}{2},|\tau|\geq 1, \tau_2>0\right\}~.
\label{fd}
\end{equation}

Since the longitudinal oscillator part is cancelled with the ghost
sector, the path-integral along this direction reduces to the summation
over zero-mode momenta. In the case of Minkowski space
the longitudinal momenta are completely decoupled from 
the transverse sector, and we can easily perform the Gaussian integral
of them (after performing the Wick rotation),
which yields the correct modular weight $\sim 1/\tau_2$.

However, in the present case, 
the transverse spectrum non-trivially depends on the longitudinal 
momenta, namely,  $p$, $\eta$ in our previous notation.
This feature is the main difficulty of calculating the partition 
function.  We  must carefully sum up over the longitudinal
momenta after evaluating the transverse conformal blocks.
Unfortunately, one can find that the naive calculation of 
the longitudinal sector leads to a divergence, even if performing
the Wick rotation. A way to avoid  this difficulty 
is to evaluate the {\em thermal\/}
partition function, which amounts to compactifying the  
Euclidean time to a circle with the circumference $\beta$ corresponding to
the inverse temperature. 

We first construct the suitable conformal blocks in the
transverse sector, and then try to evaluate the longitudinal part of the
path-integral as the thermal model.

~
\subsection{Conformal Blocks in Transverse Sector}

First of all, 
to make the problem concrete we shall assume the Gepner type 
construction \cite{Gepner}; $\cM= \cM_{k_1}\otimes \cdots \otimes \cM_{k_r}$,
where $\cM_k$ denotes the $k$-th $\cN=2$ minimal model 
($\dsp \hat{c}(\equiv c/3)= \frac{k}{k+2}$), although it is 
in principle possible to work with  more general models of rational
SCFT. The criticality condition is given as 
\begin{equation}
\sum_{i=1}^r\frac{k_i}{k_i+2}=3~.
\end{equation}
For later convenience we set
\begin{equation}
K= \mbox{L.C.M}\{k_i+2\}~,
\end{equation}
then, the possible $U(1)_R$-charge in $\cM$ sector is quantized as 
\begin{equation}
Q = \frac{a}{K}~, ~~~(a \in \bz)~.
\end{equation}

The most non-trivial part of constructing the conformal blocks in the 
transverse sector 
is taking account of the GSO projection \eqn{GSO}. As in the Gepner models,
we need the ``twists'' by the total $U(1)_R$-charge
$I_{\msc{tot},\,0}\equiv I_{H_4,\,0}+I_{\cM,\,0}$ as well as 
the projection of this charge. In this sense  we should regard 
our enhanced SUSY models as the ``orbifolds'' $(H_4 \times \cM)/\bz_K$.

The conformal blocks we want should be decomposed into the contributions 
from  (i) the (spectrally flowed) type I representations and (ii)
the type II, III representations. 
Since the spectral flows 
relate the type II and type III representations as discussed in
\cite{HS};
\begin{equation}
\cH^{(II)}_{p,\eta} \cong \cH^{(III)}_{p+1,\eta-1}~,
\label{II III}
\end{equation}
it is enough to  consider only  the flowed type II
representations. Therefore, we shall assume $0\leq \eta<1$
from here on.

The type I sector ($\eta=0$) is quite easy. 
As we already demonstrated,
the GSO condition \eqn{GSO} leads  to 
\begin{equation}
\cM/\bz_K \cong \mbox{Gepner model for $CY_3$}~.
\end{equation}
Namely, the conformal blocks in this sector is the same as those 
appearing in the Gepner model describing $CY_3$.

The type II representations  are more interesting and can 
include new sectors not appearing in the usual $CY_3$ compactifications.
Since the $U(1)_R$-charge of twisted Fock vacuum is equal to $-\eta$, 
the condition \eqn{GSO} leads to 
\begin{equation}
\eta = Q \equiv \frac{a}{K}~ (\mod ~ \bz)~.
\label{U(1) projection}
\end{equation} 
Therefore,  
the fundamental conformal blocks in the $H_4$ sector 
reduce to those of the $\bc/\bz_K$-orbifold;
 (we use the notation $y\equiv e^{2\pi i z}$, 
$q\equiv e^{2 \pi i \tau}$ from now on)
\begin{eqnarray}
&&f^{(\sNS)}_{(a,b)}(\tau, z) \df y^{-a/K}
\frac{\th_3\left(\tau, -z+\frac{a\tau+b}{K}\right)}
{\th_1\left(\tau,\frac{a\tau+b}{K}\right)}~, 
\nn
&&f^{(\stNS)}_{(a,b)}(\tau, z) \df y^{-a/K}
\frac{\th_4\left(\tau, -z+\frac{a\tau+b}{K}\right)}
{\th_1\left(\tau,\frac{a\tau+b}{K}\right)}~,   \nn
&&f^{(\sR)}_{(a,b)}(\tau, z) \df y^{-a/K}
\frac{\th_2\left(\tau, -z+\frac{a\tau+b}{K}\right)}
{\th_1\left(\tau,\frac{a\tau+b}{K}\right)}~, \nn
&&f^{(\stR)}_{(a,b)}(\tau, z) \df  y^{-a/K}
\frac{\th_1\left(\tau, -z+\frac{a\tau+b}{K}\right)}
{\th_1\left(\tau,\frac{a\tau+b}{K}\right)}~, 
\label{fab}
\end{eqnarray}
where $a,b \in \bz~, 0\leq a ,b <K, ~ (a,b)\neq (0,0)$.

In order to introduce the conformal blocks in $\cM$ sector we start with 
fixing a particular modular invariant;
\begin{equation}
Z_{\cM}(\tau,\bar{\tau})=\frac{1}{2}\sum_{I,\bar{I}}\sum_{\al}\,
N_{I,\bar{I}} F^{(\al)}_I(\tau,0)F^{(\al)}_{\bar{I}}(\bar{\tau},0)~,
\label{part M}
\end{equation} 
where $\al$ runs over the spin structures $\NS,~\tNS,~\R, ~\tR$.
$F^{(\al)}_I(\tau,z)$ are defined as the products of characters of 
the minimal models $\cM_{k_i}$;
\begin{eqnarray}
&& F_I^{(\al)}(\tau,z) = \prod_{i=1}^r\, \ch{(\al)}{l_i,m_i}(\tau,z)~,~~~
\mbox{for $\al=\NS$, $\tNS$} ~,\nn
&& F_I^{(\sR)}(\tau,z) = \prod_{i=1}^r\, \ch{(\sR)}{l_i,m_i-1}(\tau,z)~, \nn 
&& F_I^{(\stR)}(\tau,z) = (-1)^r\prod_{i=1}^r\, 
\ch{(\stR)}{l_i,m_i-1}(\tau,z)~,
\end{eqnarray}  
where $I$ denotes the collective indices 
$I \equiv \left((l_1,m_1),\ldots,  (l_r,m_r)\right)$ and 
$l_i+m_i \in 2\bz$.  
$F^{(\al)}_I(\tau,z)$ generally  possess the following modular properties
\begin{eqnarray}
&& F^{(\al)}_I(\tau+1,z) = 
e^{2\pi i \gamma(I,\al)} F^{(T\cdot \al)}_I(\tau,z)~,\\
&& F^{(\al)}_I(-\frac{1}{\tau},\frac{z}{\tau}) = 
e^{3\pi i \frac{z^2}{\tau}} \sum_{J, \beta}\, \cS_{(I,\al), (J,\beta)}
F^{(\beta)}_J(\tau,z)~,
\end{eqnarray}
where we set
\begin{eqnarray}
&& \gamma(I,\al) \df  \sum_{i=1}^r \frac{l_i(l_i+2)-m_i^2}{4(k_i+2)} + s(\al)
    ~, \nn
&& s(\al) = \left\{
\begin{array}{ll}
 -\frac{3}{8}&  ~~ (\al = \NS,~\tNS)  \\
 0& ~~ (\al=\R,~\tR) ~,
\end{array}
\right. 
\end{eqnarray}
and introduced the notation
\begin{eqnarray}
&& T\cdot\NS = \tNS~, ~~~ T\cdot\tNS = \NS~,~~~ T\cdot \R = \R~,~~~
T\cdot \tR = \tR ~.
\end{eqnarray}
For $S$-transformation, we can generally find the following form of 
modular $S$-matrix
\begin{equation}
\cS_{(I,\al),(J,\beta)} = \left(
\begin{array}{cccc}
 S_{IJ}& 0 & 0 & 0\\
  0    & 0 & e^{-i\pi Q(I)}S_{IJ}& 0 \\
  0    & e^{-i\pi Q(J)}S_{IJ} & 0 & 0 \\
  0    & 0 & 0   & -ie^{-i\pi (Q(I)+Q(J))}S_{IJ}
\end{array}
\right) ~,
\label{S-matrix}
\end{equation}
where we set (the ``total $U(1)_R$-charge'')
\begin{equation}
Q(I) \df \sum_{i=1}^r \frac{m_i}{k_i+2}~,
\end{equation}
and $S_{IJ}$ can be calculated from the knowledge of $\cN=2$ minimal model.
In the expression of \eqn{S-matrix} the rows and columns 
correspond to $\al= \NS,~\tNS,~\R,~\tR$. 
The modular invariance of \eqn{part M} requires the next condition
\begin{eqnarray}
&&  \sum_{I,\bar{I}, \al, \bar{\al}}\, 
 N_{I,\bar{I}}\delta_{\al\bar{\al}}  
  \cS_{(I,\al),(J,\beta)}
  \cS^*_{(\bar{I},\bar{\al}),(\bar{J},\bar{\beta})}
  = N_{J,\bar{J}}\delta_{\beta\bar{\beta}}~,
\label{N S} \\
&& N_{I,\bar{I}}=0~, ~~~
\mbox{unless $\gamma(I,\al)- \gamma(\bar{I},\al) \equiv 0~ (\mod~1)$}~.
\label{N T}
\end{eqnarray}
The simplest example satisfying  \eqn{N S} and \eqn{N T} is of course 
the diagonal modular invariant;
\begin{equation}
N_{I,\bar{I}}= \frac{1}{2^r}\prod_{i=1}^r \left(
 \delta_{l_i, \bar{l}_i} \delta_{m_i,\bar{m}_i}
+ \delta_{k_i-l_i\bar{l}_i} \delta_{m_i+k_i+2,\bar{m}_i}  
\right)~.
\end{equation}
(The second term is due to the field identification of minimal model.)

Now, the task we have to do is the orbifold procedure which imposes
the GSO condition \eqn{GSO}.  We must consider 
several twists by the total $U(1)_R$-charge $I_{\msc{tot},0}\equiv
I_{H_4,0}+I_{\cM,0}$ 
both along the spatial and temporal directions. 
To this aim it is convenient to introduce the ``spectral flow invariant
orbits'' as in the Gepner model cases (see, for example, \cite{EOTY}).
We first focus on the NS sector.
The actions of spectral flows with integral parameters are realized as
the procedure $z~\rightarrow~z+m\tau+n$ $(m,n \in \bz)$, and we set
\begin{equation}
F^{(\sNS)}_{I,(m,n)}(\tau,z) \df
q^{\frac{3}{2}m^2}y^{3m}F^{(\sNS)}_I(\tau,z+m\tau+n)~, ~~~ (m,n \in \bz)~.
\label{NS m n}
\end{equation}
This function possesses  the next periodicity
\begin{equation}
F^{(\sNS)}_{I,(m+Kr,n+Ks)}(\tau,z) =F^{(\sNS)}_{I,(m,n)}(\tau,z)
~, ~~~ ({}^{\forall} r,s \in \bz)~. 
\end{equation}
We thus only have to concentrate on the range $m,n \in \bz_K$
when  considering the summation with respect to the integral spectral flows.
The summation 
$\dsp \frac{1}{K}\sum_{m,n\in \bsz_K} F^{(\sNS)}_{I,(m,n)}$ 
yields the flow invariant orbits in the Gepner model for $CY_3$ \cite{EOTY} 
in which only the states with the integral $I_{\cM,0}$ charges survive.
However, since our GSO condition is now modified to \eqn{GSO},
we must construct the orbits suitably including the functions
$f^{(\sNS)}_{(a,b)}(\tau,z)$.   The condition \eqn{GSO}
needs the phase factor $e^{-2\pi i \frac{na}{K}}$ in the summation
of spectral flows. However, we must rather employ the factor
$e^{2\pi i \frac{mb-na}{K}}$ to realize good modular properties.
In this way the desired flow invariant orbits should be
\begin{eqnarray}
\cF^{(\sNS)}_{I,(a,b)}(\tau,z) &\df& \frac{1}{K}
\sum_{m,n\in \bsz_K} e^{2\pi i \frac{mb-na}{K}}
f^{(\sNS)}_{(a,b)}(\tau,z) F^{(\sNS)}_{I,(m,n)}(\tau,z) \nn
&\equiv & \frac{1}{K}
\sum_{m,n\in \bsz_K} e^{2\pi i \frac{mb-na}{K}} q^{\frac{3}{2}m^2}y^{3m}
f^{(\sNS)}_{(a,b)}(\tau,z) F^{(\sNS)}_I(\tau,z+m\tau+n)~.
\label{orbit NS}
\end{eqnarray}
By construction it is obvious that
$\cF^{(\sNS)}_{I,(a,b)}(\tau,z)\equiv 0$ unless 
$\dsp Q(I) \equiv \frac{a}{K}~ (\mod~1)$ (that is, the GSO condition
\eqn{GSO}).
For other spin structures
the flow orbits are defined with the helps of half integral spectral
flows;
\begin{eqnarray}
&& \cF^{(\stNS)}_{I,(a,b)}(\tau,z)  
\df \cF^{(\sNS)}_{I,(a,b)}(\tau, z+\frac{1}{2}) \nn
&&\hspace{1.5cm} \equiv  \frac{1}{K}
\sum_{m,n\in \bsz_K} e^{2\pi i \frac{mb-(n+\frac{1}{2})a}{K}} 
q^{\frac{3}{2}m^2}y^{3m} (-1)^m
f^{(\stNS)}_{(a,b)}(\tau,z) F^{(\sNS)}_I(\tau,z+m\tau+(n+\frac{1}{2}))~, \nn
&& \cF^{(\sR)}_{I,(a,b)}(\tau,z) 
\df q^{\frac{1}{2}}y^2\cF^{(\sNS)}_{I,(a,b)}(\tau, z+\frac{\tau}{2}) \nn
&&\hspace{1.5cm}\equiv   \frac{1}{K}
\sum_{m,n\in \bsz_K} e^{2\pi i \frac{(m+\frac{1}{2})b-na}{K}} 
q^{\frac{3}{2}(m+\frac{1}{2})^2}y^{3(m+\frac{1}{2})} 
f^{(\sR)}_{(a,b)}(\tau,z) F^{(\sNS)}_I(\tau,z+(m+\frac{1}{2})\tau+n)~, \nn
&& \cF^{(\stR)}_{I,(a,b)}(\tau,z) 
\df  q^{\frac{1}{2}}y^2\cF^{(\sNS)}_{I,(a,b)}
(\tau, z+\frac{\tau}{2}+\frac{1}{2}) \nn
&&\hspace{1.5cm} \equiv   \frac{1}{K}
\sum_{m,n\in \bsz_K} e^{2\pi i \frac{(m+\frac{1}{2})b-(n+\frac{1}{2})a}{K}} 
q^{\frac{3}{2}(m+\frac{1}{2})^2}y^{3(m+\frac{1}{2})}(-1)^m 
f^{(\stR)}_{(a,b)}(\tau,z) \nn
&& \hspace{1in} \times F^{(\sNS)}_I
(\tau,z+(m+\frac{1}{2})\tau+(n+\frac{1}{2}))~.
\label{orbit}
\end{eqnarray}

It is convenient to also introduce the conformal blocks of the $a=b=0$ sector,
which corresponds to the type I representations;
\begin{eqnarray}
\cF^{(\sNS)}_{I,(0,0)}(\tau,z) &\df& \frac{1}{K} 
\frac{1}{(2\pi)^2 2\tau_2} \frac{\th_3(\tau,z)}{\eta(\tau)^3}
\sum_{m,n\in \bsz_K} F^{(\sNS)}_{I,(m,n)}(\tau,z) \nn
&\equiv & \frac{1}{K}
\frac{1}{(2\pi)^2 2\tau_2} \frac{\th_3(\tau,z)}{\eta(\tau)^3}
\sum_{m,n\in \bsz_K} q^{\frac{3}{2}m^2}y^{3m}
F^{(\sNS)}_I(\tau,z+m\tau+n)~,
\label{orbit NS 2}
\end{eqnarray}
and also,
\begin{eqnarray}
\cF^{(\stNS)}_{I,(0,0)}(\tau,z)  
&\df& \cF^{(\sNS)}_{I,(0,0)}(\tau, z+\frac{1}{2}) ~, \nn
 \cF^{(\sR)}_{I,(0,0)}(\tau,z) 
&\df& q^{\frac{1}{2}}y^2\cF^{(\sNS)}_{I,(0,0)}(\tau, z+\frac{\tau}{2}) ~,\nn
 \cF^{(\stR)}_{I,(0,0)}(\tau,z) 
&\df & q^{\frac{1}{2}}y^2\cF^{(\sNS)}_{I,(0,0)}
(\tau, z+\frac{\tau}{2}+\frac{1}{2}) ~.
\label{orbit 2}
\end{eqnarray}
In the expression \eqn{orbit NS 2}, 
$\dsp \frac{1}{K}\sum_{m,n\in \bsz_K} F^{(\sNS)}_{I,(m,n)}(\tau,z)$
correspond to  the orbits for $CY_3$, and 
the factor $\sim 1/\tau_2$ originates from the zero-mode integral 
along the transverse plane $Z$ and  $Z^*$.

The modular properties of flow orbits $\cF^{(\al)}_{I,(a,b)}(\tau,z)$
are given by straightforward calculations;
\begin{eqnarray}
\cF^{(\al)}_{I,(a,b)}(\tau+1,z) &=& e^{2\pi i \hat{\gamma}(I,a,\al)}
\cF^{(T\cdot \al)}_{I,(a,b+a)}(\tau,z)~, \label{orbit T} \\
\cF^{(\al)}_{I,(a,b)}(-\frac{1}{\tau},\frac{z}{\tau})
&=& i e^{i\pi\frac{4z^2}{\tau}}\sum_J S_{IJ} 
\cF^{(S\cdot\beta)}_{J,(b,-a)}(\tau,z)~,
\label{orbit S}
\end{eqnarray}
where $S_{IJ}$ were  defined in \eqn{S-matrix} and 
we introduced the notation 
\begin{eqnarray}
&& S\cdot\NS = \NS~, ~~~ S\cdot\tNS = \R~,~~~ S\cdot \R = \tNS~,~~~
S\cdot \tR = \tR ~,
\end{eqnarray}
as before. $\hat{\gamma}(I,a,\al)$ is defined as 
\begin{eqnarray}
\hat{\gamma}(I,a,\al) &\df& \sum_{i=1}^r 
\frac{l_i(l_i+2)-m_i^2}{4(k_i+2)}-\frac{1}{2}Q(I)
+\frac{a}{2K}+\hat{s}(\al)~,\nn
 \hat{s}(\al) &\df& 
\left\{
\begin{array}{ll}
 -\frac{1}{2}& ~~ (\al = \NS,~\tNS) \\
 0  & ~~ (\al =\R, ~\tR) ~.
\end{array}
\right. 
\end{eqnarray}
We also remark the next spectral flow symmetry
\begin{eqnarray}
&& q^{2r^2}y^{4r} \cF^{(\al)}_{I,(a,b)}(\tau,z+r\tau+s) = 
\cF^{(\al)}_{I,(a,b)}(\tau,z)~, ~~~ ({}^{\forall} r,s \in \bz)~,
\label{orbit spectral flow}
\end{eqnarray}
which is obvious by construction.

At this stage it is quite easy to construct a modular invariant. 
We must sum up over the spin structures
both in the left and right movers independently (to impose the 
GSO projection in the usual sense).
One subtlety is the existence of redundancy
within the representations appearing in the flow invariant orbits 
$\cF^{(\al)}_{I,(a,b)}$. We need  renormalize properly  the modular
invariant coefficients $N_{I,\bar{I}}$ to avoid overcounting.
We thus do it so that the coefficient of  the ``graviton orbit'', 
which is the orbit of identity representation in  $\cM$
and resides in the (flowed) type I representations, is fixed to be 1,
according to \cite{EOTY}. We write the modular coefficients 
renormalized in this way as $\hat{N}_{I,\bar{I}}$, and finally obtain 
the following partition function
\begin{eqnarray}
Z(\tau,\bar{\tau}) = \frac{1}{4}
\sum_{a,b \in \bsz_{K}}\, 
\sum_{\al,\bar{\al}}\,\sum_{I,\bar{I}}\,
\ep(\al)\ep(\bar{\al})
\hat{N}_{I,\bar{I}}\cF^{(\al)}_{I,(a,b)}(\tau,0)
{\cF^{(\bar{\al})}_{\bar{I},(a,b)}(\tau,0)}^*~,
\label{part orbifold}
\end{eqnarray}
where we introduced the symbol
\begin{equation}
\ep(\al)=+1 ~~ \mbox{for } \al = \NS,~\tR ~, ~~~
\ep(\al)=-1 ~~ \mbox{for } \al = \tNS,~\R ~.
\end{equation} 
One can confirm straightforwardly  the modular invariance 
with the helps of  the relations \eqn{orbit T} and  \eqn{orbit S}.

However, this is {\em not\/} the desired partition function
of our enhanced SUSY model, since the transverse Hilbert space
should depend on the longitudinal momenta $p$, $\eta\equiv a/K$
as we already mentioned. In particular,  we must correctly
impose the level matching condition
\begin{equation}
L_0^{\msc{tr}}-\bar{L}_0^{\msc{tr}} \in (p+\frac{a}{K})\bz~.
\label{lm pa}
\end{equation}
The modulus integral $\dsp \int d\tau_1$ of 
\eqn{part orbifold} leads us to the level matching condition 
in the flat background
\begin{equation}
L_0^{\msc{tr}}-\bar{L}_0^{\msc{tr}} =0 ~,
\end{equation}
rather than \eqn{lm pa}. In fact, the modular invariant \eqn{part
orbifold} is no other than the partition function of
a simpler  string vacuum 
$\br^{1,1} \times \left( (\bc/\bz_K) \times \cM \right)/\bz_K$ 
(up to normalization),
where the overall denominator $\bz_K$ means 
the orbifoldization associated with the GSO projection 
as in Gepner model. 
Nevertheless, we can use $\cF^{(\al)}_{I,(a,b)}(\tau,z)$ 
as the correct building blocks of 
desired partition function we will discuss later.

Let us next argue on  the consistency of the conformal blocks 
$\cF^{(\al)}_{I,(a,b)}(\tau,z)$ with the existence of space-time SUSY.

~


\subsection{Consistency with Space-time SUSY}

It is an important consistency check of our conformal blocks 
$\cF^{(\al)}_{I,(a,b)}(\tau,z)$ \eqn{orbit NS}, \eqn{orbit},
\eqn{orbit NS 2}, \eqn{orbit 2}
to confirm the cancellation of NS and R sectors, namely,
\begin{equation}
 \sum_{\al} \ep(\al)\cF^{(\al)}_{I,(a,b)}(\tau,z) \equiv 0~,~~~
({}^{\forall} I, a, b)~.
\label{SUSY cancellation}
\end{equation}
In order to show that  this is indeed the case,  
we first note the fact that
the total conformal system $\cM\times \{Z,Z^*,\psi,\psi^*\}$
is an $\cN=2$ SCFT with $c=12$. 
The conformal blocks 
$\cF^{(\al)}_{I,(a,b)}(\tau,z)$ correspond to (reducible)
unitary representations of $c=12$ $\cN=2$ SCA and hence 
they must be decomposed into  unitary irreducible characters of 
$c=12$ $\cN=2$ SCA.
Furthermore, it is obvious by our construction that the conformal blocks
should be expanded by the ``extended characters''
with coefficients of positive integers,
which are defined by summing up of the irreducible characters
over the integral spectral flows.
More precisely, the extended characters are defined by the relation such as
\begin{equation}
\Ch{(\sNS)}{}(*;\tau,z)= \sum_{m\in \bsz} q^{2m^2}y^{4m}
\ch{(\sNS)}{}(*;\tau,z+m\tau)~,
\end{equation}
where $\ch{(\sNS)}{}(*;\tau,z)$ denotes the character of an unitary 
irreducible representation $\cN=2$ SCA with $c=12$. 
Such characters can be regarded as the ones of 
the extended superconformal algebras characteristic for the 
superstrings on $SU(n)$ holonomy manifolds (in the case of $c=3n$), 
often called ``$c=3n$ algebras''.
They  are defined by adding the spectral flow generators to
the original $\cN=2$ SCA. The most familiar example is of course 
the $c=6$ case, in which the extended algebra is no other than the 
(small) $\cN=4$ SCA with level 1 and the properties of $\cN=4$
characters as the spectral flow sum of $\cN=2$ characters 
are clarified in \cite{ET,EOTY}. For the $c=9$ case parallel analyses
are presented in \cite{Odake,EOTY},
and the explicit forms of the extended characters are given in \cite{Odake}.

For the present case of $c=12$ we can work out  the similar analysis and  
the extended characters are classified as
\begin{itemize}
 \item three continuous series of ``massive representations'', which contain
 no null states in the spectra:
(i) $\Ch{(\al)}{}(h,Q=0;\tau,z)~(h>0)$, 
(ii) $\Ch{(\al)}{}(h,Q=+1;\tau,z)~(h>1/2)$,
and (iii) $\Ch{(\al)}{}(h,Q=-1;\tau,z)~(h>1/2)$.
\item four ``massless representations'', which contain null states and 
 (anti) chiral primary states  $\dsp h=\frac{|Q|}{2}$ as the vacuum states:
(i) $\Ch{(\al)}{0}(Q=0;\tau,z)$, (ii) $\Ch{(\al)}{0}(Q=+1;\tau,z)$,
(iii) $\Ch{(\al)}{0}(Q=-1;\tau,z)$, and (iv) $\Ch{(\al)}{0}(Q=|2|;\tau,z)$.
(In the fourth case the vacuum states are doubly degenerated, $h=1,\,Q=2$
and $h=1,\,Q=-2$.)
\end{itemize}
For example, the massive character (in NS sector) 
$\Ch{(\sNS)}{}(h,Q=0;\tau,z)$ is calculated as
\begin{equation}
\Ch{(\sNS)}{}(h,Q=0;\tau,z) = q^{h-3/8}\Th{0}{3/2}(\tau,2z)
\frac{\th_3(\tau,z)}{\eta(\tau)^3}~.
\end{equation}
The detailed analysis and the explicit forms 
of all the other characters are summarized in Appendix B. 
Among other things, we can show the identities
\begin{equation}
\sum_{\al} \ep(\al)\Ch{(\al)}{(0)}(*;\tau,0) \equiv 0~, 
\end{equation}
for all the extended characters. 
This fact directly proves the supersymmetric cancellation of our
conformal blocks \eqn{SUSY cancellation}. 
Since the discussion here is quite general, we can apply this result
to arbitrary unitary $\cN=2$ SCFTs with $c=12$, which is relevant 
for arbitrary compactifications of $SU(n)$ holonomy manifolds.

~

\subsection{Thermal Partition Functions of Enhanced SUSY Models}

Now, let us compute the longitudinal part of partition function
as a thermal model as we already declared. We first recall 
the spectrum of longitudinal momentum $p^+(\equiv \cF)$;
\begin{equation}
p^+ = p+\frac{a}{K}\equiv \frac{pK+a}{K}~,
~~~(p\in \bz_{\geq 0},~ 0\leq a <K) ~.
\label{p+}
\end{equation}
We emphasize that this spectrum
is discretized by the GSO condition \eqn{U(1) projection} 
and the rationality of $\cM$ sector. 
The level matching condition of transverse sector 
is written as \eqn{lm pa}, which is derived from
the condition \eqn{J}
$p^--\bar{p}^- (\equiv \cJ-\bar{\cJ})=h \in \bz$.
On the other hand, the modular invariance also requires 
$L_0^{\msc{tr}}-\bar{L}_0^{\msc{tr}}\in \bz$.
Therefore we shall assume the next level matching condition stronger 
than \eqn{lm pa};
\begin{equation}
L_0^{\msc{tr}}-\bar{L}_0^{\msc{tr}} \in (pK+a)\bz~.
\label{new lm}
\end{equation}

It is remarkable that the spectra \eqn{p+} and \eqn{new lm}
are formally equivalent to  those of 
DLCQ string theory \cite{DLCQ} with the compactification radius $R=K$.
The thermal partition function of DLCQ string theory has been 
calculated in \cite{Semenoff} and we can make use of their result.
Let us now present a very short review of it.

We first consider the bosonic string case for simplicity.
In the Wick rotated space-time 
$\dsp X^{\pm}\equiv \frac{1}{\sqrt{2}}(X^1\pm i X_E^0)$,
the DLCQ string theory ($X^-\sim X^-+2\pi R$) is 
described by the identification
\begin{equation}
X^0_E\sim X^0_E+\sqrt{2}\pi R i~,~~~ X^1\sim X^1+\sqrt{2}\pi R~,
\label{E DLCQ}
\end{equation}
and the thermal compactification is defined as
\begin{equation}
X^0_E\sim X^0_E+\beta~,
\end{equation}
where $\beta$ denotes the inverse temperature.
When calculating the Polyakov path-integral, the longitudinal 
oscillator part is cancelled out with the ghost sector. 
The calculation of zero-mode part reduces to summing up of the
classical action
over the ``instantons'' with various winding numbers $m,n,r,s$;
\begin{eqnarray}
&& X^0_E(w+2\pi,\bar{w}+2\pi)= X^0_E(w,\bar{w})+\beta m +\sqrt{2}\pi R ir~,\nn
&& X^0_E(w+2\pi\tau,\bar{w}+2\pi\bar{\tau})= 
X^0_E(w,\bar{w})+\beta n +\sqrt{2}\pi R is~,\nn
&& X^1(w+2\pi,\bar{w}+2\pi)=X^1(w,\bar{w})+\sqrt{2}\pi Rr~,\nn
&& X^1(w+2\pi\tau,\bar{w}+2\pi\bar{\tau})=X^1(w,\bar{w})+\sqrt{2}\pi Rs~.
\label{instanton}
\end{eqnarray}
Note that the sector of $m=n=0$, which corresponds to  the vacuum energy in 
the zero temperature limit, yields a divergent contribution. 
We should subtract it and hence assume $(m,n)\neq (0,0)$ in the summation.
The summation over $r$, $s$ can be now easily carried out and gives  
a periodic delta function on moduli space of torus.
We hence obtain 
\begin{eqnarray}
 \int\cD\lb X^+,X^-\rb
\cD\lb \mbox{gh}\rb\, e^{-S_{\msc{L}}-S_{\msc{gh}}} &=& \nu
\sum_{m,n,p,q}\,e^{-\frac{\beta^2|m\tau-n|^2}{8\pi\tau_2}}
\delta^{(2)}\left((m\nu+ip)\tau-(n\nu+iq)\right) \nn
&\equiv& \frac{1}{\tau_2} \rho(\tau,\bar{\tau})~,
\label{partition l}
\end{eqnarray}
where we set $\dsp \nu = \frac{\sqrt{2}\beta R}{8\pi}$.
Clearly $\rho(\tau,\bar{\tau})$ is modular invariant;
\begin{eqnarray}
\rho(\tau+1,\bar{\tau}+1)=\rho(\tau,\bar{\tau})~, ~~~
\rho(-1/\tau,-1/\bar{\tau})=\rho(\tau,\bar{\tau})~,
\end{eqnarray}
and thus 
\begin{equation}
Z_{\msc{1-loop}} = \int_{\cF} \frac{d^2\tau}{\tau_2^2}\rho(\tau,\bar{\tau}) 
Z^{\msc{tr}}(\tau,\bar{\tau})
\end{equation}
is the correct form of one-loop partition function of string theory.

For the type II superstring case we only have to modify the function
$\rho(\tau,\bar{\tau})$ so as to include suitably the spin structures of
world-sheet fermions \cite{AW,Semenoff}. More precisely, we obtain
\begin{eqnarray}
&&\rho^{(\al,\bar{\al})}(\tau,\bar{\tau}) = \nu
\sum_{m,n,p,q}\,\tau_2 e^{-\frac{\beta^2|m\tau-n|^2}{8\pi\tau_2}}
\kappa(\al;m,n)\kappa(\bar{\al};m,n) \nn
&& \hspace{3cm} \times
\delta^{(2)}\left((m\nu+ip)\tau-(n\nu+iq)\right) ~,
\label{rho IIA}\\
&& \kappa(\NS;m,n)\df 1 ~,~~~\kappa(\tNS;m,n)\df (-1)^m~,\nn
&& \kappa(\R;m,n)\df (-1)^n~,~~~\kappa(\tR;m,n)\df (-1)^{m+n}~.
\end{eqnarray}
The phase factors $\kappa(\al;m,n)$ are most easily
understood by recalling the correct boundary conditions
in the thermal field theory of point particles (for the $m=0$ cases)
and further taking account of the consistency with modular invariance.


Now, let us return to the present problem. It seems enough to
simply replace $R$ with $K$ in the above result \eqn{rho IIA}.
However, because the boundary conditions of transverse string coordinates
$Z,Z^*,\psi,$ and $\psi^*$ are related to the longitudinal momentum 
$\dsp p^+ = p+\frac{a}{K}$, 
we need a slight modification in order to recover the correct 
level matching condition.
We can make a simple guess  that the following decomposition
of \eqn{rho IIA} works as the correct modification;
\begin{eqnarray}
\rho^{(\al,\bar{\al})}(\tau,\bar{\tau})&=& 
\sum_{a,b\in \bsz_{K}}\rho^{(\al,\bar{\al})}_{(a,b)}(\tau,\bar{\tau})~,\\
\rho^{(\al,\bar{\al})}_{(a,b)}(\tau,\bar{\tau}) &\df&  \nu
\sum_{m,n,p,q}\,\tau_2 e^{-\frac{\beta^2|m\tau-n|^2}{8\pi\tau_2}}
\kappa(\al;m,n)\kappa(\bar{\al};m,n) \nn
&& \times
\delta^{(2)}\left((m\nu+i(pK+a))\tau-(n\nu+i(qK-b))\right) ~.
\label{rho ab}
\end{eqnarray}
The functions
$\rho_{(a,b)}^{(\al,\bar{\al})}(\tau,\bar{\tau})$ have 
the periodicity
\begin{equation}
\rho^{(\al,\bar{\al})}_{(a+pK,b+qK)}(\tau,\bar{\tau})
=\rho^{(\al,\bar{\al})}_{(a,b)}(\tau,\bar{\tau})~,
\end{equation}
and the expected modular properties
\begin{eqnarray}
\rho_{(a,b)}^{(\al,\bar{\al})}(\tau+1,\bar{\tau}+1)
=\rho_{(a,b+a)}^{(T\cdot\al,T\cdot\bar{\al})}(\tau,\bar{\tau})~,~~~
\rho_{(a,b)}^{(\al,\bar{\al})}(-1/\tau,-1/\bar{\tau})
=\rho_{(b,-a)}^{(S\cdot\al,S\cdot\bar{\al})}(\tau,\bar{\tau})~.
\end{eqnarray}
Therefore we propose the following partition function as 
the correct thermal partition function, 
of which validity is confirmed just below;
\begin{eqnarray}
Z_{\msc{1-loop}}
&=&\int_{\cF}\,\frac{d^2\tau}{\tau_2^2}\,\frac{1}{4}
\sum_{a,b}\sum_{\al,\bar{\al}}
\sum_{I,\bar{I}}\rho^{(\al,\bar{\al})}_{(a,b)}(\tau,\bar{\tau})
\ep(\al)\ep(\bar{\al})\hat{N}_{I,\bar{I}}
\cF^{(\al)}_{I,(a,b)}(\tau,0)
\cF^{(\bar{\al})}_{\bar{I},(a,b)}(\tau,0)^* \nn
&\equiv&\frac{1}{4}
\sum_{m,n,p,q}\sum_{a,b}\sum_{\al,\bar{\al}}\sum_{I,\bar{I}}
\,\frac{\nu e^{-\frac{\beta^2|m\tau-n|^2}
       {8\pi\tau_2}}}{m^2\nu^2+(pK+a)^2}
\frac{1}{\tau_2}\kappa(\al;m,n)\kappa(\bar{\al};m,n)\ep(\al)\ep(\bar{\al})
\nn
&& \hspace{3cm} \times \hat{N}_{I,\bar{I}}
\cF^{(\al)}_{I,(a,b)}(\tau,0)\cF^{(\bar{\al})}_{\bar{I},(a,b)}(\tau,0)^*~,
\label{partition function}
\end{eqnarray}
where we set 
\begin{equation}
\tau= \frac{n\nu+i(qK-b)}{m\nu+i(pK+a)} ~,
\label{modulus 1}
\end{equation}
in the last line. Recall that the parameter $\nu$ is defined 
as $\nu = \sqrt{2}\beta K/8\pi$ (since the DLCQ radius is now equal to $K$).
The summation with respect to $m,n,p,q$
should be taken over the range such that $\tau \in \cF$.

It is obvious by construction that the integrand of \eqn{partition
function} is modular invariant, and thus it has a consistent form 
of one-loop partition function of string theory. 
It is enough to confirm that  the level matching condition \eqn{lm pa} or
\eqn{new lm}  is recovered  in order to check the validity of 
this partition function. 
For this purpose it is easiest to make use of the following observation 
as presented in \cite{Polchinski,Semenoff}.
We first note that \eqn{partition function}
has the form such as 
\begin{equation}
Z_{\msc{1-loop}}
=\sum_{m,n}\int_{\cF}\frac{d^2\tau}{\tau_2^2} f_{(m,n)}(\tau,\bar{\tau})~,
\end{equation}
where $m$, $n$ denote the winding numbers defined in \eqn{instanton}.
$f_{(m,n)}(\tau,\bar{\tau})$ manifestly possesses the next modular 
property;
\begin{eqnarray}
&& f_{(m',n')}(\tau',\bar{\tau}') = f_{(m,n)}(\tau,\bar{\tau}) ~, \nn
&& \tau' = \frac{a\tau+b}{c\tau+d}~, ~~~{}^{\forall} A=
\left(
\begin{array}{cc}
 a&b \\
 c&d
\end{array}
\right) \in SL(2;\bz) ~,\nn
&& (m',n') = (m,n)A^{-1}~.
\end{eqnarray}
We can always find out a modular transformation such that $m'=0$, $n'<0$
for arbitrary $(m,n) \neq (0,0)$. 
Therefore, we can take a different 
``gauge choice'' $m=0$, $n<0$ and  the modulus integral 
should be carried out in a larger domain
\begin{equation}
\cF' \equiv \left\{\tau\in \bc~;~ |\tau_1| \leq \frac{1}{2},~ \tau_2 >0
\right\}~,
\end{equation} 
rather than $\cF$.

In this way we can rewrite \eqn{partition function} as a simpler form;
\begin{eqnarray}
Z_{\msc{1-loop}} 
&=& \frac{1}{4}
\sum_{n,p,q}\sum_{a,b}\sum_{\al,\bar{\al}}\sum_{I,\bar{I}}\,
\frac{e^{-\frac{\beta^2n^2}{8\pi\tau_2}}}{n(pK+a)}
\kappa(\al;0,n)\kappa(\bar{\al};0,n)\ep(\al)\ep(\bar{\al})
\nn
&& \hspace{2cm} \times \hat{N}_{I,\bar{I}}
\cF^{(\al)}_{I,(a,b)}(\tau,0)\cF^{(\bar{\al})}_{\bar{I},(a,b)}(\tau,0)^*~,
\label{partition function 2}
\end{eqnarray}
where the integers $n,p,q,a,b$ run  over the range 
such that $\dsp \tau \equiv \frac{qK-b+in\nu}{pK+a} \in \cF'$.

Observing this expression, we can confirm that the summation over $q$
with  $\dsp \tau_1=\frac{qK-b}{pK+a}$ imposes the correct level matching
condition \eqn{lm pa} (and necessarily \eqn{new lm}). 
We thus conclude that \eqn{partition function} is the 
thermal partition function of $(H_4\times \cM)/\bz_K$ model we want. 

~

Finally we make a few comments:

~

\noindent{\bf 1.} 
Although the conformal blocks $\cF^{(\al)}_{I,(a,b)}$ are supersymmetric
as we already discussed, the space-time SUSY in the thermal model is 
completely broken. In fact, the partition function 
\eqn{partition function} (or \eqn{partition function 2}) does not vanish
because of the existence of extra phase factor $\kappa(\al;m,n)$.

~

\noindent{\bf 2.}
The free energy in the second quantized 
free string theory with finite temperature is computed as 
\begin{eqnarray}
F&=&\frac{1}{\beta}
\tr \left\lb (-1)^{\msc{\bf F}}
\ln \left(1-(-1)^{\msc{\bf F}}e^{-\beta p^0}\right)\right\rb \nn
&\equiv& -\sum_{n=1}^{\infty}\frac{1}{\beta n}
\tr \left\lb (-1)^{(n+1)\msc{\bf F}}e^{-\beta n p^0}\right\rb ~,
\label{free energy}
\end{eqnarray}
where $\mbox{\bf F}$ denotes the space-time fermion number (mod 2)
and $\dsp p^0 \equiv \frac{1}{\sqrt{2}}(p^+ - p^-)$ 
is the space-time energy operator. The trace should be 
taken over the single particle physical Hilbert space on which 
the on-shell condition and the level matching condition are imposed. 
In the present case we can rewrite by means of the on-shell condition
as
\begin{equation}
p^0 =  \frac{1}{\sqrt{2}}\left(\frac{pK+a}{K}+ \frac{K}{2(pK+a)}
(L_0^{\msc{tr}}+\bar{L}_0^{\msc{tr}}-1)\right)~.
\end{equation}
With the help of this equality  and by observing 
the expression \eqn{partition function 2}, it is not difficult
to show the equality 
\begin{equation}
Z_{\msc{1-loop}} = - \beta  F~.
\end{equation} 
This fact provides  the correct relation between the free energy 
in the second quantized theory and the one-loop partition function
in the first quantized thermal string.  We thus believe the consistency of  
our result \eqn{partition function} (or \eqn{partition function 2}).

~

\subsection{Thermal Partition Functions of $H_4 \times CY_3$ Models}

Finally, let us discuss the one-loop partition functions of 
string vacua $H_4 \times CY_3$ defined by \eqn{GSO 0} in order 
to accomplish  our study. We again compute it as the thermal model. 

First of all, the conformal blocks in $\cM$ sector are
calculated independently of the $H_4$ sector;
\begin{eqnarray}
&& \cG^{(\sNS)}_I(\tau,z) \df \frac{1}{K}\sum_{m,n\in \bsz_K}\,
  F_{I,\,(m,n)}^{(\sNS)}(\tau,z)~,
\end{eqnarray}
and the blocks for other spin structures are defined by the
half integral spectral flows as before. 
They are the flow invariant orbits describing
the $\sigma$ model on $CY_3$ as already mentioned.

We calculate the partition function with discretizing the twist
parameter $\eta$ as $\eta = a/N$ ($N$ is an integer independent 
of $K$ previously defined), and then consider the large $N$ limit.
Before taking the large $N$ limit, we can obtain the thermal partition
function by the similar calculations. However, we need a slight modification
here.  Since we must take the range $p \in 2\bz$ due to the 
locality of space-time supercharges \eqn{SUSY charge 0},
we have to now use both of the type II and III representations. 
Therefore, the range of summation of $a,b$ should be $a,b \in \bz_{2N}$
rather than $a,b \in \bz_{N}$.
We then obtain  
\begin{eqnarray}
&& Z_{\msc{1-loop}} = 
\int_{\cF}\,\frac{d^2\tau}{\tau_2^2}\,\frac{1}{4}
\sum_{a,b \in \bsz_{2N}}\sum_{\al,\bar{\al}}
\rho^{(\al,\bar{\al})}_{(a,b)}(\tau,\bar{\tau})
\ep(\al)\ep(\bar{\al}) f^{(\al)}_{(a,b)}(\tau,0)
f^{(\bar{\al})}_{(a,b)}(\tau,0)^*  \nn
&& \hspace{3cm} \times \sum_{I,\bar{I}}
\hat{N}_{I,\bar{I}}
\cG^{(\al)}_{I}(\tau,0)
\cG^{(\bar{\al})}_{\bar{I}}(\tau,0)^* ~,
\label{partition function 3}
\end{eqnarray}
where $f^{(\al)}_{(a,b)}(\tau,0)$ ($(a,b)\neq (0,0)$) are defined in
\eqn{fab} with the integer $K$ replaced with $N$ (for the type II and 
III representations), and $f^{(\al)}_{(0,0)}(\tau,z)$ are 
defined as  (for the type I representations)
\begin{eqnarray}
&& f^{(\sNS)}_{(0,0)}(\tau,z) \df 
\frac{1}{(2\pi)^2 2\tau_2} \frac{\th_3(\tau,z)}{\eta(\tau)^3}~,~~~
f^{(\stNS)}_{(0,0)}(\tau,z) \df 
\frac{1}{(2\pi)^2 2\tau_2} \frac{\th_4(\tau,z)}{\eta(\tau)^3}~, \nn
&& f^{(\sR)}_{(0,0)}(\tau,z) \df 
\frac{1}{(2\pi)^2 2\tau_2} \frac{\th_2(\tau,z)}{\eta(\tau)^3}~,~~~
f^{(\stR)}_{(0,0)}(\tau,z) \df 
\frac{1}{(2\pi)^2 2\tau_2} \frac{\th_1(\tau,z)}{\eta(\tau)^3}~.
\label{fab 2}
\end{eqnarray} 
$\rho^{(\al,\bar{\al})}_{(a,b)}(\tau, \bar{\tau})$ is also defined in 
\eqn{rho ab} again with the replacement of $K$ by $N$, and $p$, $q$
by $2p$, $2q$.

Under the large $N$ limit, it is easy to see that 
the contribution from type I representations vanishes, 
and that of type II and III representations can be evaluated 
by replacing the sum over $a$, $b$ with the integral; 
\begin{eqnarray}
\frac{1}{4N^2}\sum_{a,b} F(a/N, b/N)~\longrightarrow~ 
\int_{-1}^1 du\int_{-1}^1 dv \,F(u,v) ~.
\end{eqnarray}
Obviously we must include the divergent volume factor 
$V\equiv \sqrt{2}\pi \beta N$.  We finally  obtain 
\begin{eqnarray}
&& Z_{\msc{1-loop}} 
= \frac{V}{8\pi^2}
\int_{\cF}\,\frac{d^2\tau}{\tau_2^2}\, 
\int_{-1}^{1}du\int_{-1}^{1} dv\, 
\sum_{\al,\bar{\al}} \sum_{m,n,p,q}\, 
\tau_2 e^{-\frac{\beta^2|m\tau-n|^2}{8\pi\tau_2}} 
 \kappa(\al;m,n)\kappa(\bar{\al};m,n) \nn
&& \hspace{2cm} \times
\delta^{(2)}\left(\left(\frac{\sqrt{2}\beta}{8\pi}m+i(2p+u)\right)\tau
-\left(\frac{\sqrt{2}\beta}{8\pi}n+i(2q-v)\right)
\right) \nn
&& \hspace{2cm} \times
\ep(\al)\ep(\bar{\al}) g^{(\al)}_{(u,v)}(\tau,0)
g^{(\bar{\al})}_{(u,v)}(\tau,0)^*  \, 
\sum_{I,\bar{I}} \hat{N}_{I,\bar{I}}
\cG^{(\al)}_{I}(\tau,0)
\cG^{(\bar{\al})}_{\bar{I}}(\tau,0)^* ~,
\label{partition function 4}
\end{eqnarray}
where  we set $g^{(\al)}_{(u,v)}(\tau,z) \df f^{(\al)}_{(a,b)}(\tau,z)$
with the identifications $u\equiv a/N$, $v\equiv b/N$,  $(a,b)\neq (0,0)$.
The integrand of modulus integral in \eqn{partition function 4}
is manifestly modular invariant.

We note that the transverse conformal blocks are not cancelled 
in contrast to the enhanced SUSY model $(H_4\times \cM)/\bz_K$;
\begin{equation}
\sum_{\al} \ep(\al)g^{(\al)}_{(u,v)}(\tau,z) \cG^{(\al)}_I(\tau,z) 
\not\equiv 0~, ~~~ ({}^{\forall} (u,v)\not\in \bz\times\bz,~ {}^{\forall} I)~.
\end{equation}
This result is not a contradiction, because the SUSY charges 
\eqn{SUSY charge 0} do not commute with the light-cone Hamiltonian
$H_{\msc{l.c.}}\equiv -(\cJ+\bar{\cJ})$. However, as we commented before,
if one uses the light-cone Hamiltonian (or the transverse Virasoro
operators) of the type given in \cite{RT2,RT} to define the conformal 
blocks, one can find that the supersymmetric cancellation occurs
(only for the left-mover) in the same way as that of superstring vacua
$\br^{3,1}\times CY_3$.

We finally comment on the zero-temperature limit $\beta \rightarrow
\infty$. This is equal to the vacuum energy and captured only by  
the $m=n=0$ sector.   
Although we have subtracted this sector in defining the thermal
partition function, it is interesting to set formally $m=n=0$
in the expression of \eqn{partition function 4}. 
The integrations of the parameters $u$, $v$ 
are easily carried out because of the delta function factor, and 
$g_{(u,v)}^{(\al)}$ reduces to 
\begin{equation}
\sim \tau_2^{-1}\times \mbox{infinite volume factor} \times 
f^{(\al)}_{(0,0)}~,
\end{equation}
where $f^{(\al)}_{(0,0)}$ is defined in \eqn{fab 2}.
Therefore, the vacuum energy becomes the partition function of 
$\br^{3,1}\times CY_3$ 
 (up to an infinite volume factor)\footnote
     {The same result may be also derived from the fact that 
      the vacuum energy in the zero-temperature does not depend 
      on the DLCQ radius (equal to $N$ in the present case) as discussed
      in \cite{Semenoff}.}.
This result is likely to be consistent with the observation 
given in \cite{KK,RT2,TT}.

~

\section{Discussions}

In this paper we have explored  superstring vacua  
constructed from the conformal theory $H_4 \times \cM$. 
The choice of GSO projection is a key ingredient.
The simplest choice \eqn{GSO 0} gives the background $H_4\times CY_3$. 
These string vacua have a manifest geometric interpretation 
and have unbroken SUSY  (4 supercharges) that is  
consistent with the analysis of Killing spinor in supergravity. 
In fact, it is shown that the half of maximal SUSY 
with a particular chirality associated with one of 
the light-cone coordinate are left unbroken (see, for example, 
\cite{BKO,BFHP}).
Such unbroken SUSY corresponds to the supercharges made up of 
the spin fields of the types $S^{+**}$ in our context 
and is described by \eqn{SUSY charge 0} more explicitly.

On the other hand, the enhanced SUSY vacua are defined by the GSO
condition \eqn{GSO} and have 8 supercharges, namely,  the maximal 
SUSY in the Calabi-Yau compactification of type II string.
Typical models of such string vacua can be constructed from 
the string theories on $AdS_3 \times S^1 \times \cM'$ by taking
the Penrose limit, and the number of unbroken SUSY is consistent
with this fact. Although the bulk physics seems to be
the same as in the first case $H_4 \times CY_3$, 
which corresponds to the sectors of 
free strings described by the spectrally flowed type I representations,
these backgrounds do not have a naive geometrical interpretation. 
In particular, one cannot regard the target space as 
a simple direct product because of the orbifoldization
associated with the GSO projection \eqn{GSO}.

Nevertheless, we can get an intuitive insight for the reason why
we can obtain the enhanced SUSY by observing the simple example
$H_6 \times T^4$. This is obtained by taking the Penrose limit of $AdS_3
\times S^3 \times T^4$ and is a special case of our enhanced SUSY vacua
as we mentioned in section 3.
This background is described by the $\sigma$ model \cite{RT}
\begin{equation}
 L = \partial u \bar{\partial} v + \cF_{ij} x^i \partial u \bar{\partial} x^j
   + \partial x^i \bar{\partial} x^i ~,
\end{equation}
where $i,j=1,\ldots,8$ and the NSNS-flux is given 
by $\cF_{ij} = f
\epsilon_{ij} (i,j = 1,2)$ (for the $AdS_3$ direction)
and $\cF_{kl} = f \epsilon_{kl} (k,l
=3,4)$ (for the $S^3$ direction). 
By using the notation of the gamma matrices in Appendix A of
\cite{HS}, one can find that the relevant condition of Killing spinor
reduces to
\begin{equation}
  \Gamma^{+0} (\Gamma^{+1} \Gamma^{-1} - \Gamma^{-2} \Gamma^{+2} ) 
 \epsilon = 0 ~.
\label{killing3}
\end{equation}
Therefore, except for the Killing spinors satisfying $\Gamma^{+0} \epsilon =
0$, there are 8 Killing spinors which satisfy 
$(\Gamma^{+1} \Gamma^{-1} - \Gamma^{-2} \Gamma^{+2} ) 
\epsilon = 0 $\footnote
   {The same conclusion has been obtained in the recent paper
    \cite{Bena}.}.
In the notation of \cite{HS}  the former Killing spinors
leads to the 16 supercharges 
\begin{eqnarray}
&& \cQ^{++a} = \oint S^{+++aa} e^{i X^+} ~,~~~
 \cQ^{--a} = \oint S^{+--aa} e^{- i X^+} ~, \nn
&& \cB_0^{+a} = \oint S^{+-+(-a)a} ~,~~~
 \cB_0^{-a} = \oint S^{++-(-a)a} ~,
\end{eqnarray}
as well as the counterparts of right movers,
which corresponds to \eqn{SUSY charge 1}.
The latter Killing spinors correspond to the 8 extra supercharges
\begin{equation}
 \cQ^{-+a} = \oint S^{--+aa} ~,~~~
 \cQ^{+-a} = \oint S^{-+-aa} ~,
\end{equation}
which corresponds to \eqn{SUSY charge 2}.
These extra supercharges generate the super transformation
which preserves the light-cone Hamiltonian and give rise to 
the cancellation between the NS sectors and R sectors.

The above analysis gives us  an important suggestion that the existence of 
extra supercharges reflects the ``cancellation''  of  
NSNS-flux essentially captured in the equation (\ref{killing3}).
Therefore, it may be plausible to expect that our enhanced SUSY 
vacua are described geometrically by Calabi-Yau spaces with suitable
NSNS-flux which cancels that of $H_4$ background. 
In fact, we remark here the similarity of 
the construction of our string vacua $(H_4 \times \cM)/\bz_K$
to the Gepner models.  In that models the orbifoldization with respect to
the $U(1)_R$-charge ensures the locality of supercharges and also
implies the existence of non-vanishing NSNS-flux.
Further study about precise geometrical interpretation of our 
enhanced SUSY vacua could be significant, and  
it is  quite interesting to discuss the relation 
to the several works about the classification of supergravities 
on pp-waves which possess extra SUSY \cite{extra,Bena}.

As for the  one-loop partition functions, we have evaluated them as the
thermal models.  As a byproduct we have proved the following statement
at the character level: Every $\cN=2$ unitary SCFTs
with $c=12$ exhibit  the cancellation of space-time SUSY under 
the integrality condition of $U(1)_R$-charge.
This is the most general statement of SUSY cancellation applicable
to arbitrary compactifications on $SU(n)$-holonomy manifolds
including non-compact models \cite{singular CY} as well as 
the Gepner models.

When calculating the thermal partition functions,
the  similarity to the DLCQ string played 
an important role. Also in the $AdS_3$ string such similarity
appears and was clarified in \cite{HHS} at the level of free field
representation.
The features as DLCQ theory in these string vacua may be profound 
for the possibility of approach of Matrix string theory \cite{MST}
to the studies of pp-wave physics. Attempts along this direction have
been given in the recent papers \cite{Gopakumar,Bonelli,Verlinde}. 
It may be also interesting to compare our result with 
the thermal partition function in the $AdS_3$ string, which was 
calculated in \cite{MOS}.

~


\section*{Acknowledgement}
\indent

We  would like to thank T. Eguchi and  T. Takayanagi for valuable discussions.
The work of Y. S. is supported in part by a
Grant-in-Aid for the Encouragement of Young Scientists 
($\sharp 13740144$) from the Japanese Ministry of Education, 
Culture, Sports, Science and Technology.

~

~


\section*{Appendix A ~ Notations}
\setcounter{equation}{0}
\def\theequation{A.\arabic{equation}}
\indent

In this appendix we summarize the conventions used in this paper.
We set $q\equiv e^{2\pi i \tau}$ and  $y\equiv e^{2\pi i z}$.

~

\noindent
{\bf 1. Theta functions} 
 \begin{eqnarray}
&& \th_1(\tau,z) =i\sum_{n=-\infty}^{\infty}(-1)^n q^{(n-1/2)^2/2} y^{n-1/2}
  \equiv 2 \sin(\pi z)q^{1/8}\prod_{m=1}^{\infty}
    (1-q^m)(1-yq^m)(1-y^{-1}q^m), \nn
&&  \th_2(\tau,z)=\sum_{n=-\infty}^{\infty} q^{(n-1/2)^2/2} y^{n-1/2}
  \equiv 2 \cos(\pi z)q^{1/8}\prod_{m=1}^{\infty}
    (1-q^m)(1+yq^m)(1+y^{-1}q^m), \nn
&& \th_3(\tau,z)=\sum_{n=-\infty}^{\infty} q^{n^2/2} y^{n}
  \equiv \prod_{m=1}^{\infty}
    (1-q^m)(1+yq^{m-1/2})(1+y^{-1}q^{m-1/2}), \nn
&& \th_4(\tau,z)=\sum_{n=-\infty}^{\infty}(-1)^n q^{n^2/2} y^{n}
  \equiv \prod_{m=1}^{\infty}
    (1-q^m)(1-yq^{m-1/2})(1-y^{-1}q^{m-1/2}) ,
\end{eqnarray}
 \begin{eqnarray}
 \Th{m}{k}(\tau,z)&=&\sum_{n=-\infty}^{\infty}
 q^{k(n+\frac{m}{2k})^2}y^{k(n+\frac{m}{2k})} ~,\\
 \tTh{m}{k}(\tau,z)&=&\sum_{n=-\infty}^{\infty} (-1)^n
 q^{k(n+\frac{m}{2k})^2}y^{k(n+\frac{m}{2k})} ~.
 \end{eqnarray}
 We also use the standard convention of $\eta$-function;
 \begin{equation}
 \eta(\tau)=q^{1/24}\prod_{n=1}^{\infty}(1-q^n)~.
 \end{equation}

~


\noindent
{\bf 2. Character formulas of $\cN=2$ minimal model}

The character formulas of the $k$-th $\cN=2$ minimal model 
($\dsp c= \frac{3k}{k+2}$) are given as follows;
\begin{eqnarray}
\ch{(\sNS)}{l,m}(\tau,z)&=&\chi^{l,0}_m(\tau,z)+\chi^{l,2}_m(\tau,z) ~,\nn
\ch{(\stNS)}{l,m}(\tau,z)&=&\chi^{l,0}_m(\tau,z)-\chi^{l,2}_m(\tau,z) ~,\nn
\ch{(\sR)}{l,m}(\tau,z)&=&\chi^{l,1}_m(\tau,z)+\chi^{l,3}_m(\tau,z)~,\nn
\ch{(\stR)}{l,m}(\tau,z)&=&\chi^{l,1}_m(\tau,z)-\chi^{l,3}_m(\tau,z)~,
\label{minimal character}
\end{eqnarray}
where $\chi^{l,s}_m(\tau,z)$ is defined by
\begin{equation}
\chi_m^{l,s}(\tau,z)=\sum_{r\in \bsz_k}c^{(k)}_{l,m-s+4r}(\tau)
\Th{2m+(k+2)(-s+4r)}{2k(k+2)}(\tau,z/(k+2))~.
\label{branching}
\end{equation}
In the  expression  of \eqn{branching} we assume $l+m+s\equiv 0~(\mod~2)$,
and $c^{(k)}_{l,m}$ denotes the level $k$ string function of $SU(2)$,
which is defined by the well-known relation
\begin{equation}
\chi^{(k)}_l(\tau, z)\left(\equiv \frac{\Th{l+1}{k+2}-\Th{-l-1}{k+2}}
                        {\Th{1}{2}-\Th{-1}{2}} (\tau,z) \right)
= \sum_{m\in \bsz_{2k}}c^{(k)}_{l,m}(\tau)\Th{m}{k}(\tau,z)~.
\end{equation}
By definition $\chi^{l,s}_{m}$ has the following periodicity
\begin{equation}
\chi^{l,s}_{m+2(k+2)}=\chi^{l,s+4}_{m}=\chi^{k-l,s+2}_{m+k+2}=\chi^{l,s}_m~.
\end{equation}


~

\section*{Appendix B ~ Character Formulas of 
``$c=12$ Extended Superconformal Algebra'' }
\setcounter{equation}{0}
\def\theequation{B.\arabic{equation}}
\indent

We set  
\begin{eqnarray}
f^{(\sNS)}(\tau, z) &\df & q^{1/8}\frac{\th_3(\tau,z)}{\eta(\tau)^3} \equiv
\frac{\prod_{n=1}^{\infty}(1+yq^{n-1/2})(1+y^{-1}q^{n-1/2})}
   {\prod_{n=1}^{\infty}(1-q^n)^2}~
\end{eqnarray}
for convenience.
We first focus on the NS sector and later consider the other spin
structures with the help of half integral spectral flows.

~

\noindent{\bf 1. Massive representations}

The ``massive representation'' of $c=12$ $\cN=2$ SCA 
is quite easy, since no null states are included in the spectrum.
Such unitary representation is characterized by the conformal weight $h$
and $U(1)_R$-charge $Q$ of vacuum state $(h>\frac{Q}{2})$,
and the character formula is simply written as 
\begin{eqnarray}
\ch{(\sNS)}{}(h,Q;\tau,z) &=& q^{h-\frac{1}{2}}y^Qf^{(\sNS)}(\tau,z)~.
\label{massive}
\end{eqnarray}
The character formula of corresponding representation in 
the ``$c=12$ algebra'' is constructed by summing
over the integral spectral flows;
\begin{eqnarray}
\Ch{(\sNS)}{}(h,Q;\tau,z) &=& \sum_{m\in\bsz}q^{2m^2}y^{4m}
\ch{(\sNS)}{}(h,Q;\tau,z+m\tau)~.
\label{massive 2}
\end{eqnarray}
If we assume the integral $U(1)_R$-charges\footnote
     {In the cases of $c=6$ algebra ($\cN=4$ SCA with level 1)
     and $c=9$ algebra \cite{Odake} the integrality of $U(1)_R$-charges
      simply originates from the unitarity of representations.
     However, in our case of $c=12$ algebra the situation is more subtle.
     In any case we shall here assume the integrality of $U(1)_R$-charges,
     which is enough for our purpose since our conformal blocks 
      $\cF^{(\al)}_{I,(a,b)}(\tau,z)$ should have this property by 
      construction.}, the possible massive 
representations of $c=12$ algebra are classified into the next 
three continuous series;
(i) $h>0$, $Q=0$, (ii) $h>1/2$, $Q=1$, and 
(iii) $h>1/2$, $Q=-1$.
The character formulas \eqn{massive 2} can be expressed in terms of 
the level $3/2$ theta functions for the each case;
\begin{eqnarray}
\Ch{(\sNS)}{}(h,Q=0;\tau,z) &=& q^{h-1/2}\,\Th{0}{3/2}(\tau,2z)\,
f^{(\sNS)}(\tau,z)~,   \nn 
\Ch{(\sNS)}{}(h,Q=\pm 1;\tau,z) &=& q^{h-2/3}\,\Th{\pm 1}{3/2}(\tau,2z)\,
f^{(\sNS)}(\tau,z)~.
\label{massive 3}
\end{eqnarray}


~

\noindent
{\bf 2. Massless representations}

The ``massless representations'' of $c=12$ algebra
can be constructed by the spectral flows
based on the following degenerate representations;
(i) $h=Q=0$,
(ii) $h=1/2$, $Q=1$, 
(iii) $h=1/2$, $Q=-1$, and
(iv) $h=1$, $Q=\pm 2$.
(In the fourth case the vacuum state doubly degenerates in the sense of 
$c=12$ algebra, $Q=2$ and $Q=-2$.)
For each of these cases the $\cN=2$ characters are given in 
\cite{Dobrev} based on the data of Kac determinant formula 
for $\cN=2$ SCA \cite{BFK}; 
\begin{eqnarray}
\ch{(\sNS)}{0}(Q=0;\tau,z) &=& q^{-1/2}
\frac{1-q}{(1+yq^{1/2})(1+y^{-1}q^{1/2})} f^{(\sNS)}(\tau,z)~, \\
\ch{(\sNS)}{0}(Q;\tau,z) &=& q^{-1/2} 
\frac{q^{\frac{|Q|}{2}}y^{Q}}{1+y^{\msc{sign}(Q)}q^{1/2}}f^{(\sNS)}(\tau,z)~,
~~~(Q=\pm 1,\pm 2)~.
\end{eqnarray}
The massless characters of $c=12$ algebra are again obtained 
by summing up over the integral spectral flows as in \eqn{massive 2}.
The results are written as
\begin{eqnarray}
\Ch{(\sNS)}{0}(Q=0;\tau,z) &=& 
q^{-1/2}\sum_{m\in\bsz}\frac{(1-q)q^{\frac{3}{2}m^2+m-\frac{1}{2}}y^{3m+1}}
{(1+yq^{m+1/2})(1+yq^{m-1/2})} f^{(\sNS)}(\tau,z) \nn 
&\equiv &
q^{-1/2}\,\sum_{m\in\bsz}\,
\frac{yq^{m-1/2}-1}{1+yq^{m-1/2}}\,q^{\frac{3}{2}m^2}y^{3m}f^{(\sNS)}(\tau,z)
\nonumber \\
&& ~~~-q^{-1/6}\,\left(\Th{1}{3/2}(\tau,2z)-\Th{-1}{3/2}(\tau,2z)\right)\,
f^{(\sNS)}(\tau,z) ~,  \\
\Ch{(\sNS)}{0}(Q=\pm 1;\tau,z) &=&q^{-1/2}\, \sum_{m\in\bsz}\,
\frac{1}{1+y^{\pm 1}q^{m+1/2}}\,q^{\frac{3}{2}m^2+m+\frac{1}{2}}
y^{\pm (3m+1)}f^{(\sNS)}(\tau,z) ~,    \\
\Ch{(\sNS)}{0}(Q=|2|;\tau,z) &=&q^{-1/2}\, \sum_{m\in\bsz}\,
\frac{1}{1+yq^{m+1/2}}\,q^{\frac{3}{2}m^2+2m+1}
y^{3m+2}f^{(\sNS)}(\tau,z) ~.
\end{eqnarray}

The following identities are useful;
\begin{eqnarray}
\Ch{(\sNS)}{0}(Q=1;\tau,z) &=&\Ch{(\sNS)}{0}(Q=-1;\tau,-z)  \nonumber \\
&=&  \Ch{(\sNS)}{0}(Q=-1;\tau,z) \nonumber\\
&& ~~~+q^{-1/6}\,\left(\Th{1}{3/2}(\tau,2z)
  -\Th{-1}{3/2}(\tau,2z)\right)\,f^{(\sNS)}(\tau,z)~, \\
\Ch{(\sNS)}{0}(Q=0;\tau,z) &=&\Ch{(\sNS)}{0}(Q=0;\tau,-z)~, \\
\Ch{(\sNS)}{0}(Q=|2|;\tau,z) &=&\Ch{(\sNS)}{0}(Q=|2|;\tau,-z)~. 
\end{eqnarray} 
We also remark the following relations between massless and massive
characters;
\begin{eqnarray}
&&q^h\left(\Ch{(\sNS)}{0}(Q=0;\tau,z)+\Ch{(\sNS)}{0}(Q=1;\tau,z)
+\Ch{(\sNS)}{0}(Q=-1;\tau,z)\right)
\nonumber \\
&& \hspace{3cm} = \Ch{(\sNS)}{}(h,Q=0;\tau,z)  ~,\\
&& q^{h-1/2}\left(
\Ch{(\sNS)}{0}(Q=\pm 1;\tau,z)+\Ch{(\sNS)}{0}(Q=|2|;\tau,z)\right) \nn
&& \hspace{3cm}= \Ch{(\sNS)}{}(h,Q=\pm 1;\tau,z)~. 
\end{eqnarray}
In other words, the massive characters can be decomposed into the massless
characters at the threshold $h\,\rightarrow\, 0 (1/2) $. 
This aspect is completely parallel to the $c=6$ case 
\cite{ET} and the $c=9$ case \cite{Odake}.

The most important property of these extended characters 
in our discussion is the cancellation due to the space-time supersymmetry. 
In order to observe it manifestly we define the characters 
with the other spin structures by the half integral spectral flows;
\begin{eqnarray}
\Ch{(\stNS)}{(0)}(*;\tau,z) &\df& \Ch{(\sNS)}{(0)}(*;\tau,z+\frac{1}{2})~, \nn
\Ch{(\sR)}{(0)}(*;\tau,z) &\df&q^{1/2}y^2 
\Ch{(\sNS)}{(0)}(*;\tau,z+\frac{\tau}{2}) ~, \nn
\Ch{(\stR)}{(0)}(*;\tau,z) &\df&q^{1/2}y^2 
\Ch{(\sNS)}{(0)}(*;\tau,z+\frac{\tau}{2}+\frac{1}{2})~.
\end{eqnarray}   
The twisted Ramond characters  $\Ch{(\stR)}{(0)}(*;\tau,0)$ 
are no other than the Witten indices.
We can easily find 
\begin{eqnarray}
&& \Ch{(\stR)}{}(h,Q;\tau,0) =0~,
\end{eqnarray}
for the arbitrary massive representations and 
\begin{eqnarray}
&& \Ch{(\stR)}{0}(Q=0;\tau,0)=2~, ~~~\Ch{(\stR)}{0}(Q=\pm 1;\tau,0)=-1~,\nn
&& \Ch{(\stR)}{0}(Q=|2|;\tau,0)=1~,
\end{eqnarray}
for the massless representations.

Then, the identities of supersymmetry are written as
\begin{eqnarray}
&& \sum_{\al}\ep(\al)\Ch{(\al)}{}(h,Q=0;\tau,z) \equiv 0~,~~~ 
({}^{\forall} h) 
\label{SUSY identity 1} \\
&& \sum_{\al}\ep(\al)\left(\Ch{(\al)}{}(h,Q=+1;\tau,z)
+\Ch{(\al)}{}(h,Q=-1;\tau,z)\right) \equiv 0~,~~~ ({}^{\forall} h)
\label{SUSY identity 2} \\
&& \sum_{\al}\ep(\al)\Ch{(\al)}{0}(Q=0, |2|;\tau,z) \equiv 0~, 
\label{SUSY identity 3} \\
&& \sum_{\al}\ep(\al)\left(\Ch{(\al)}{0}(Q=+1;\tau,z)
+\Ch{(\al)}{0}(Q=-1;\tau,z)\right) \equiv 0~,
\label{SUSY identity 4} 
\end{eqnarray} 
where $\ep(\NS)=\ep(\R)=+1$ and $\ep(\tNS)=\ep(\tR)=-1$ as before.
The identities \eqn{SUSY identity 1} and \eqn{SUSY identity 2}
reduce to the known theta function identities which are directly
proved by the product formula;
\begin{eqnarray}
&&\Th{0}{3/2}(\tau,2z)\th_3(\tau,z)-\tTh{0}{3/2}(\tau,2z)\th_4(\tau,z)
\nonumber \\
&& \hspace{3cm}
-\Th{3/2}{3/2}(\tau,2z)\th_2(\tau,z)+i\tTh{3/2}{3/2}(\tau,2z)\th_1(\tau,z)
\equiv 0 ~,   \label{SU3 holonomy} \\
&&  \left(\Th{1}{3/2}+\Th{-1}{3/2}\right)(\tau,2z) \th_3(\tau,z)
+\left(\tTh{1}{3/2}+\tTh{-1}{3/2}\right) (\tau,2z) \th_4(\tau,z)
\nonumber \\
&& \hspace{1cm} 
-\left(\Th{1/2}{3/2}+\Th{-1/2}{3/2}\right)(\tau,2z) \th_2(\tau,z)
-i\left(\tTh{1/2}{3/2}-\tTh{-1/2}{3/2}\right)(\tau,2z) \th_1(\tau,z)
\equiv 0~. \nonumber \\
\label{SU3 holonomy 2}
\end{eqnarray}
For the massless representations it seems difficult 
to analytically prove \eqn{SUSY identity 3} and 
\eqn{SUSY identity 4} unfortunately. 
However, we have directly confirmed
these identities by MAPLE in lower orders in $q$, $y$, $y^{-1}$
and believe their correctness.

Finally we  note that all the characters $\Ch{(\al)}{(0)}(*;\tau,z)$
have a symmetry  under the integral spectral flows, which 
precisely means
\begin{equation}
 q^{2r^2}y^{4r} \Ch{(\al)}{(0)}(*;\tau,z+r\tau+s) =
\Ch{(\al)}{(0)}(*;\tau,z)~,~~~({}^{\forall} r,s \in \bz)~.
\end{equation}
This property is obvious by construction and consistent with
the fact that the conformal blocks $\cF^{(\al)}_{I,(a,b)}(\tau,z)$
can be expanded by these characters.


\end{document}